\documentclass[reprint,twocolumn,pre,showpacs,amsmath,amssymb,aps]{revtex4-1}

\usepackage{natbib}
\usepackage{graphicx,color,rotating}
\usepackage[latin1]{inputenc}
\usepackage{textcomp}
\usepackage{dcolumn}
\usepackage{bm}     
\usepackage{upgreek}
\usepackage{subdepth}  			
\usepackage[latin1]{inputenc}

\usepackage{array}
\newcolumntype{L}[1]{>{\raggedright\let\newline\\\arraybackslash\hspace{0pt}}m{#1}}
\newcolumntype{C}[1]{>{\centering\let\newline\\\arraybackslash\hspace{0pt}}m{#1}}
\newcolumntype{R}[1]{>{\raggedleft\let\newline\\\arraybackslash\hspace{0pt}}m{#1}}

\usepackage{hyperref}					
\usepackage{xcolor}						
\hypersetup{			     			
    colorlinks,
    linkcolor={blue!80!black},
    citecolor={blue!80!black},
    urlcolor={blue!80!black}
}
\newcommand{\notop}{{{}_{}}}

\newcommand{\mr}[1]{\ensuremath{\mathrm{#1}}}
\renewcommand{\vec}[1]{\bm{#1}}
\newcommand{\ee}{\mathrm{e}}
\newcommand{\ii}{\mathrm{i}}
\newcommand{\dm}{\mathrm{d}}

\newcommand{\avr}[1]{\big\langle #1 \big\rangle}

\DeclareMathOperator{\re}{Re}

\newcommand{\iot}{{\ii\omega t}}

\newcommand{\pp}{\partial}

\newcommand{\nablabf}{\boldsymbol{\nabla}}

\newcommand{\een}{\vec{e}}

\newcommand{\FFFrad}{\vec{F}^\mathrm{rad}}
\newcommand{\FFFradbar}{\bar{\vec{F}}^\mathrm{rad}}
\newcommand{\Frad}{F^{\mathrm{rad}}}
\newcommand{\Fradnorm}{F^{\mathrm{rad}}_\mr{norm}}
\newcommand{\Fradbar}{\bar{F}^{\mathrm{rad}}}

\newcommand{\fnn}{f^{{}}}
\newcommand{\fA}{\fnn_A}
\newcommand{\fB}{\fnn_B}

\newcommand{\Hsi}{{H^\notop_\mr{si}}}	
\newcommand{\Hpy}{{H^\notop_\mr{py}}}	
\newcommand{\Hsl}{{H^\notop_\mr{sl}}}	
\newcommand{\Hsltop}{{H^{{\mr{top}}}_\mr{sl}}}
\newcommand{\Hslbot}{{H^{{\mr{bot}}}_\mr{sl}}}

\newcommand{\Hfl}{H^\notop_\mr{fl}}

\newcommand{\Lfl}{L^\notop_\mr{fl}}
\newcommand{\Lsl}{L^\notop_\mr{sl}}

\newcommand{\nnn}{\vec{n}}

\newcommand{\uuu}{\vec{u}}

\newcommand{\vvv}{\vec{v}}

\newcommand{\Wsl}{{W^\notop_\mr{sl}}}
\newcommand{\Wfl}{{W^\notop_\mr{fl}}}

\newcommand{\Zsl}{{Z^\notop_\mr{sl}}}
\newcommand{\Zfl}{{Z^\notop_\mr{fl}}}		

\newcommand{\zerovec}{\boldsymbol{0}}

\newcommand{\calA}{\mathcal{A}}

\newcommand{\calE}{\mathcal{E}}
\newcommand{\calEacfl}{\calE^\mr{fl}_\mr{ac}}
\newcommand{\calEacsl}{\calE^\mr{sl}_\mr{ac}}

\newcommand{\cfl}{c_\mr{fl}}
\newcommand{\cflsqr}{c^2_\mr{fl}}

\newcommand{\Eac}{E_\mathrm{ac}}

\newcommand{\Eacfl}{E_\mr{ac}^\mr{fl}}
\newcommand{\Eacsl}{E_\mr{ac}^\mr{sl}}

\newcommand{\kapPS}{\kappa_\mathrm{ps}}

\newcommand{\kapfl}{\kappa_\mr{fl}}

\newcommand{\Gamsl}{\Gamma_\mathrm{sl}}

\newcommand{\Omegafl}{{\Omega_\mr{fl}}}
\newcommand{\Omegasl}{{\Omega_\mr{sl}}}

\newcommand{\rhofl}{\rho_\mr{fl}}
\newcommand{\rhosl}{\rho_\mr{sl}}

\newcommand{\pnorm}{p_\mathrm{norm}}

\newcommand{\rhoPS}{\rho_\mathrm{ps}}

\newcommand{\SICel}{^\circ\!\textrm{C}}

\newcommand{\SIMHz}{\textrm{MHz}}

\newcommand{\SIkg}{\textrm{kg}}

\newcommand{\SIm}{\textrm{m}}

\newcommand{\SImm}{\textrm{mm}}
\newcommand{\SImum}{\textrm{\textmu{}m}}

\newcommand{\SIPa}{\textrm{Pa}}

\newcommand{\SIMPa}{\textrm{MPa}}

\newcommand{\SIs}{\textrm{s}}

\newcommand{\beq}[1]{\begin{equation} \eqlab{#1}}
\newcommand{\eeq}{\end{equation}}
\newcommand{\bsub}{\begin{subequations}}
\newcommand{\esub}{\end{subequations}}
\def\bal#1\eal{\begin{align}#1\end{align}}
\def\bsubal#1\esubal{\bsub \begin{align}#1\end{align} \esub}
\newcommand{\nn}{\nonumber}
\newcommand{\eqlab}[1]{\label{eq:#1}}
\renewcommand{\eqref}[1]{Eq.~(\ref{eq:#1})}
\newcommand{\eqnoref}[1]{(\ref{eq:#1})}

\newcommand{\eqsref}[2]{Eqs.~(\ref{eq:#1}) and~(\ref{eq:#2})}
\newcommand{\eqsnoref}[2]{(\ref{eq:#1}) and~(\ref{eq:#2})}

\newcommand{\figref}[1]{Fig.~\ref{fig:#1}}
\newcommand{\fignoref}[1]{\ref{fig:#1}}
\newcommand{\figsref}[2]{Figs.~\ref{fig:#1} and~\ref{fig:#2}}
\newcommand{\figlab}[1]{\label{fig:#1}}

\newcommand{\secref}[1]{Section~\ref{sec:#1}}

\newcommand{\seclab}[1]{\label{sec:#1}}
\newcommand{\tabref}[1]{Table~\ref{tab:#1}}
\newcommand{\tabsref}[2]{Tables~\ref{tab:#1} and~\ref{tab:#2}}
\newcommand{\tablab}[1]{\label{tab:#1}}

\newcommand{\sigmabf}{\bm{\sigma}}

\newcommand{\sigmabfsl}{\bm{\sigma}^{{}}_\mr{sl}}

\newcommand{\cL}{c_\mathrm{L}}
\newcommand{\cT}{c_\mathrm{T}}

\newcommand{\cTsqr}{c^2_\mathrm{T}}

\begin{document}

\title{Whole-system ultrasound resonances as the basis\\ for acoustophoresis in all-polymer microfluidic devices}

\author{Rayisa P. Moiseyenko}
\email{raymoi@fysik.dtu.dk}
\affiliation{Department of Physics, Technical University of Denmark, DTU Physics Building 309, DK-2800 Kongens Lyngby, Denmark}

\author{Henrik Bruus}
\email{bruus@fysik.dtu.dk}
\affiliation{Department of Physics, Technical University of Denmark, DTU Physics Building 309, DK-2800 Kongens Lyngby, Denmark}

\date{13 September 2018}

\begin{abstract}
Using a previously well-tested numerical model, we demonstrate theoretically that good acousto\-phoresis can be obtained in a microchannel embedded in an acoustically soft, all-polymer chip, by excitation of whole-system ultrasound resonances. In contrast to conventional techniques based on a standing bulk acoustic wave inside a liquid-filled microchannel embedded in an elastic, acoustically hard material, such as glass or silicon, the proposed whole-system resonance does not need a high acoustic contrast between the liquid and surrounding solid. Instead, it relies on the very high acoustic contrast between the solid and the surrounding air. In  microchannels of usual dimensions, we demonstrate the existence of whole-system resonances in an all-polymer device, which support acoustophoresis of a quality fully comparable to that of a conventional hard-walled system. Our results open up for using cheap and easily processable polymers in a controlled manner to design and fabricate microfluidic devices for single-use acoustophoresis.
\end{abstract}


\maketitle

\section{Introduction}

A steadily increasing number of papers report successful applications of ultrasound-based microscale acoustofluidic devices in biology, environmental and forensic sciences, and clinical diagnostics \cite{Lenshof2012a, Gedge2012, Sackmann2014, Laurell2014, Antfolk2017}. Examples include cell synchronization \cite{Thevoz2010}, enrichment of prostate cancer cells in blood \cite{Augustsson2012}, high-throughput cytometry and multiple-cell handling \cite{Zmijan2015, Ohlin2015}, single-cell patterning and manipulation~\cite{Collins2015, Guo2016}, size-independent sorting of cells \cite{Augustsson2016}, and rapid sepsis diagnostics by detection of bacteria in blood \cite{Ohlsson2016}. Acoustics have also been used for non-contact microfluidic trapping and particle enrichment \cite{Hammarstrom2014a}, massively parallel force microscopy on biomolecules \cite{Sitters2015}, as well as acoustic tweezing \cite{Drinkwater2016, Collins2016, Lim2016, Baresch2016}.

In all applications, an appropriate magnitude of the acoustic forces is reached by resonant actuation of ultrasound waves, using one of two basic methods. One method relies on bulk acoustic waves (BAW), see \figref{BasicWaves}(a), for which resonant modes are built up in liquid-filled acoustic resonators, say microchannels or microcavities, embedded in acoustically hard material such as silicon, glass and/or metal. For this method to work, it is crucial that the acoustic contrast between the liquid and the surrounding solid is sufficiently large, typically around 10 in terms of the acoustic impedance ratio. The other method relies on surface acoustic waves (SAW), \figref{BasicWaves}(b), that are resonantly excited by using appropriately spaced interdigitated metallic transducer electrodes positioned on the surface of a piezoelectric substrate. For this method to work, it is crucial that the piezoelectric coupling constant of the substrate is sufficiently strong, and thus a popular choice is lithium niobate substrates, well-known from conventional electromechanical filters in microwave technology. Both methods are actively being used in contemporary acoustofluidics as is evident from the following examples published in the literature the past two years. BAW devices have been used for cell focusing in simple and inexpensive aluminum devices \cite{Gautam2018}, for binary particle separation in droplet microfluidics \cite{Fornell2018}, for hematocrit determination \cite{Petersson2018}, for enrichment of tumor cells from blood \cite{Magnusson2017}, and for manipulation of \textit{C. elegans} \cite{Ahmed2016, Zhou2017}, while SAW devices have been used for nanoparticle separation \cite{Sehgal2017, Wu2017}, for self-aligned particle focusing and patterning \cite{Collins2018}, for enhanced cell sorting \cite{Ung2017}, and for in-droplet microparticle separation \cite{Park2017}.

\begin{figure}[!t]
\centering
\includegraphics[width=\columnwidth]{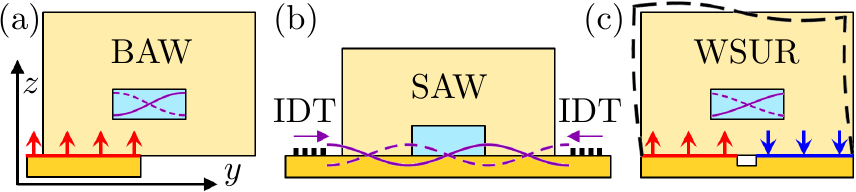}
\caption{\figlab{BasicWaves} Sketches of the conventional and proposed acousto\-phoresis techniques for a liquid-filled microchannel (blue) embedded in an elastic solid (beige). (a) Bulk acoustic waves (BAW) exited by a single piezo actuator (orange). (b) Surface acoustic waves (SAW) excited by interdigital transducers (IDT, black) on a piezoelectric substrate (orange). (c) The proposed whole-system ultrasound resonances (WSUR) excited by two piezo actuators excited in anti-phase (red and blue arrows). Magenta and dashed black curves represent pressure and displacement fields.}
\end{figure}

Currently, the acoustofluidic devices with the highest throughput are of the BAW type \cite{Antfolk2017}. However, a limiting factor for fully exploiting the application potential of such devices, is the cost of the glass or silicon components, used because of their high acoustic contrast relative to water. This limitation is especially severe for applications intended for point-of-care clinical use \cite{Sackmann2014}, where the acoustic separation unit must be a single-use consumable. This important problem could be overcome by using all-polymer microfluidic devices, if they could be made compatible with ultrasound acoustics, because that would allow for cheap conventional volume fabrication \cite{Kim2016b}. However, partly due to the lack of good theoretical understanding, it has proven difficult to make good all-polymer BAW-devices for acoustofluidics. But a few results for such devices have been published, including focusing of polymer beads \cite{Harris2012, Mueller2013, Gonzalez2015, Yang2017}, lipids \cite{Harris2012}, and red blood cells \cite{Mueller2013, Savage2017}, as well as blood-bacteria separation \cite{Silva2017} and purification of lymphocytes \cite{Lissandrello2018}.

The main goal of this work, is to provide a theoretical framework for designing all-polymer microfluidic devices capable of successful acoustophoretic applications. The structure of the paper is as follows: In \secref{GeomMaterials}, we present the basic device geometry and the material properties. In \secref{Theory}, we give a short overview of the theory of linear acoustics of the solid and the fluid, the acoustic radiation force on suspended tracer microparticles, and the numerical implementation of the model. The results of the numerical simulation for a simplified model in two dimensions (2D) of a conventional silicon-glass system are shown and analyzed in \secref{Results2D_hard}. They serve as a baseline for the 2D simulation results of an all-polymer device presented in \secref{Results2D_PMMA}, where the principle of whole-system ultrasound resonances (WSUR), see \figref{BasicWaves}(c), is established as a method to identify specific resonances suitable for successful acoustophoresis in this acoustically soft system. In \secref{Results3D}, we present the results of a more realistic model in three dimensions (3D) of the all-polymer device, and discuss them in relation to preliminary experimental results. Finally, in \secref{Conclusion}, we summarize and discuss the obtained results.

\section{Geometry and materials}
\seclab{GeomMaterials}

\begin{figure}[!b]
\centering
\includegraphics[width=\columnwidth]{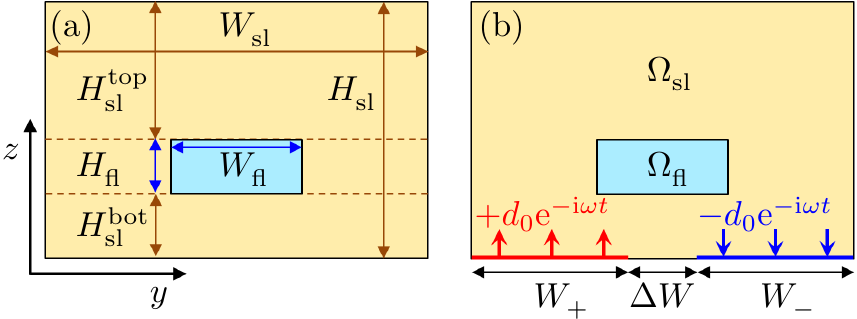}
\caption{\figlab{DeviceGeom} (a) The vertical cross section in the $y$-$z$ plane of the long, straight device. The elastic solid (beige, $\Omegasl$) has the outer widths $\Wsl$ and height $\Hsl$, while the fluid channel (light blue, $\Omegafl$) has the width $\Wfl = 377~\SImum$ and height $\Hfl = 157~\SImum$.  (b) A sketch of the anti-symmetric actuation $\uuu(W_\pm) = \pm d_0\:\ee^{-\iot}\:\nnn$ of the time-harmonic displacement at the actuation regions $W_+$ (red) and $W_-$ (blue) separated by the gap $\Delta W$. }
\end{figure}

Generic, mm-long, straight channels that are placed along the horizontal $x$ axis and have a constant rectangular cross section in the vertical $y$-$z$ plane, have been intensively studied in the literature both theoretically \cite{Muller2012, Muller2014, Muller2015, Bach2018} and experimentally \cite{Barnkob2010, Augustsson2011, Barnkob2012a, Muller2013}. This design is thus an obvious choice for our analysis, and following Refs.~\cite{Barnkob2010, Augustsson2011, Barnkob2012a, Muller2013}, the fluid domain $\Omegafl$ in the vertical $y$-$z$ plane is taken to be a rectangle of width $\Wfl = 377~\SImum$ and height $\Hfl = 157~\SImum$. The fluid domain is embedded in an elastic solid $\Omegasl$ defined by a larger rectangle of width $\Wsl$ and height $\Hsl$, see \figref{DeviceGeom}, of values to be specified below.

\begin{table}[t]
\centering
\caption{\tablab{material_param_sl} Material parameters at $25~\SICel$ for selected solids. The specific acoustic impedance is $\Zsl  = \sqrt{\rhosl C_{11}}$, and for isotropic materials $C_{12} = C_{11} - 2 C_{44}$ (PMMA and Pyrex).}
\begin{ruledtabular}
\begin{tabular}{lccccc}
 Parameter &  Symbol  & PMMA & Pyrex   & Si & Unit \\
 & & \cite{Sutherland1972, Sutherland1978, Carlson2003, Mott2008, Adler2014, Kasarova2015}
 & \cite{Corning_Pyrex} & \cite{Hopcroft2010} & \\ \hline
 Mass density    & $\rhosl$     & 1190   & 2230   & 2329   & $\SIkg\:\SIm^{-3}$  \rule{0mm}{1.1em} \\
 Elastic modulus & $C_{11}$    & 8.567  & 69.72  & 165.7  & GPa         \\
 Elastic modulus & $C_{44}$    & 1.429  & 26.15  & 79.6   & GPa         \\
 Elastic modulus & $C_{12}$    & 5.710  & 17.43  & 63.9   & GPa         \\
 Damping coeff. & $\Gamsl$      & 0.0040 & 0.0004 & 0.0004 & 1           \\
 Spec.~impedance & $\Zsl$ & 3.19 & 12.47 & 19.64 & $\SIMPa\:\SIs\:\SIm^{-1}$
\end{tabular}
\end{ruledtabular}
\end{table}

As sketched in \figref{BasicWaves}(c), the ultrasound actuation is modeled by the time-harmonic and spatially anti-symmetric displacement condition applied to the bottom surface of the solid with a frequency in the low MHz range and a fixed amplitude $d_0 \simeq 0.1$~nm, which corresponds to typical experimental values. This specific actuation is chosen to better generate the conventional anti-symmetric standing pressure half-wave in the horizontal direction, which focuses suspended particles in the vertical $x$-$z$ plane along the center $x$ axis of the channel. It is straightforward to extent this simple actuation model by adding an actual piezoelectric material to the solid and driving it by imposing a suitable ac voltage.

Material wise, the elastic solid in our model is taken to be the transparent thermoplastic polymethyl methacrylate (PMMA), also known as acrylic glass, Plexiglas or Lucite. The acoustofluidic response of this acoustically soft polymer is contrasted with that of the conventional acoustically hard devices made of a silicon base and a borosilicate glass lid (Pyrex) \cite{Barnkob2010, Augustsson2011, Barnkob2012a, Muller2013}. The values of the material parameters at ambient temperature used in the elastodynamic model of the solids are shown in \tabref{material_param_sl}. Note that in the case of PMMA we have used  representative average values based on Refs.~\cite{Sutherland1972, Sutherland1978, Carlson2003, Mott2008, Adler2014, Kasarova2015}. Material parameters at $25~\SICel$ for water and 10-$\SImum$-diameter polystyrene tracer particles are listed in \tabref{material_param_fl}.

\begin{table}[t]
\centering
\caption{\tablab{material_param_fl} Material parameters at $25~\SICel$  for water and 10-$\SImum$-diameter polystyrene tracer particles. Note: $\Zfl = \rhofl \cfl$.}
\begin{ruledtabular}
\begin{tabular}{lcrc}
Parameter &  Symbol  & Value & Unit \\ \hline
\textit{Water}: &    &    &  \rule{0mm}{1.1em} \\
Mass density \cite{Muller2014}     & $\rhofl$  & 997.05  & kg~m$^{-3}$ \\
Compressibility, isentr. \cite{Muller2014}    & $\kapfl$ & 447.7  & TPa$^{-1}$  \\
Speed of sound \cite{Muller2014}  & $\cfl$ & 1496.7  & m~s$^{-1}$ \\
Damping coefficient \cite{Hahn2015}     & $\Gamma_\mr{fl}$ & $0.004$ &  1 \vspace*{0.5mm} \\
Spec.~acoustic impedance & $\Zfl$   & 1.49 & $\SIMPa\:\SIs\:\SIm^{-1}$ \\[1mm]
\multicolumn{4}{l}{\textit{Polystyrene particles in water}:} \rule{0mm}{1.1em} \\
Mass density \cite{Hartmann1972, Karlsen2015}    & $\rhoPS$  & 1050  & kg~m$^{-3}$ \\
Compressibility \cite{Hartmann1972, Karlsen2015}  & $\kapPS$ & 238  & TPa$^{-1}$  \\
Monopole coefficient \cite{Karlsen2015} & $f_0$ & 0.468 & 1 \\
Dipole coefficient \cite{Karlsen2015}  & $f_1$ & 0.034 & 1
\end{tabular}
\end{ruledtabular}
\end{table}

\section{Theory}
\seclab{Theory}

The physical model of the acoustophoretic device consists of a fluid-filled microchannel channel embedded in an elastic solid. The piezoelectric transducer, which in reality drives the ultrasound waves in the system, is replaced by a simplifying  oscillating displacement condition on part of the outer surface of the solid. We thus need the governing equation for the linear ultrasound acoustics of the solid and the fluid, as well as of the nonlinear acoustics describing the acoustophoretic forces on microparticles suspended in the fluid. We restrict our analysis to the time-periodic and isentropic case at ambient temperature, thus disregarding transient behavior \cite{Muller2015} and thermal effects \cite{Muller2014, Karlsen2015}.

\subsection{Linear acoustics of the solid and the fluid}

We use standard, weakly damped, linear elastodynamics \cite{Landau1986} with the specific coupling to microscale acoutofluidics as formulated by Ley and Bruus \cite{Ley2017} for isotropic solids, and its extension to cubic crystals by Dual and Schwarz \cite{Dual2012}. The dynamics of the solid of density $\rhosl$ is modeled by the elastic displacement $\uuu$ and stress $\sigmabf$, both with a time-harmonic oscillation given by the phase factor $\ee^{-\iot}$, where $\omega = 2\pi f$ is the angular frequency and $f$ is the frequency. Because the governing equations are linear, this temporal phase factor is divided out in the following, leaving only spatially dependent amplitude for all field. When time dependence is needed, the phase factor $\ee^{-\iot}$ is simply reintroduced in the fields. The linear constitutive stress-strain relation is given in terms of the elastic moduli $C_{ik}$ in the Voigt representation, and the governing equation for the displacement field $\uuu$ at angular frequency $\omega$ become
 \bsub
 \eqlab{gov_eq_solid}
 \bal
 & -\rhosl \omega^2 \left(1+ \ii \Gamsl \right)^2 \uuu
 = \nablabf \cdot \sigmabf,\; \text{ with }\\[2mm]
 \eqlab{stress-strain-Voigt}
 &\left( \begin{array}{c}
 \!\!\sigma^{{}}_{xx}\!\!\! \\[1mm]
 \!\!\sigma^{{}}_{yy}\!\!\! \\[1mm]
 \!\!\sigma^{{}}_{zz}\!\!\! \\[1mm]
 \!\!\sigma^{{}}_{yz}\!\!\! \\[1mm]
 \!\!\sigma^{{}}_{xz}\!\!\! \\[1mm]
 \!\!\sigma^{{}}_{xy}\!\!\! \end{array} \right)
 =
 \left( \begin{array}{cccccc}
 \!\!C^{{}}_{11}\!\! & C^{{}}_{12}\!\! & C^{{}}_{12}\!\! & 0 & 0 & 0\!\! \\[1mm]
 \!\!C^{{}}_{12}\!\! & C^{{}}_{11}\!\! & C^{{}}_{12}\!\! & 0 & 0 & 0\!\! \\[1mm]
 \!\!C^{{}}_{12}\!\! & C^{{}}_{12}\!\! & C^{{}}_{11}\!\! & 0 & 0 & 0\!\! \\[1mm]
 \!\!0 & 0 & 0  & \!C^{{}}_{44}\!\! & 0 & 0\!\! \\[1mm]
 \!\!0 & 0 & 0  & 0 & \!C^{{}}_{44}\!\! & 0\!\! \\[1mm]
 \!\!0 & 0 & 0  & 0 & 0 & \!C^{{}}_{44}\!\!
 \end{array} \right) \!\!
  \left( \begin{array}{c}
 \!\!\pp_x u_x\!\! \\[1mm]
 \!\!\pp_y u_y\!\! \\[1mm]
 \!\!\pp_z u_z\!\! \\[1mm]
 \!\!\!\pp_y u_z\!+\!\pp_z u_y\!\!\! \\[1mm]
 \!\!\!\pp_x u_z\!+\!\pp_z u_x\!\!\! \\[1mm]
 \!\!\!\pp_x u_y\!+\!\pp_y u_x\!\!\! \end{array} \right)\! ,
 \eal
 \esub
where a cubic crystal have three independent elastic moduli $C_{11}$, $C_{12}$, and $C_{44}$, while an isotropic material has two due to the constraint $C_{12} = C_{11} - 2C_{44}$. In the latter case, the material is therefore characterized by the longitudinal and transverse sound speeds $\cL = \sqrt{C_{11}/\rhosl}$ and $\cT = \sqrt{C_{44}/\rhosl}$, respectively. The acoustic energy density $\Eacsl$ in the solid domain is the sum of the kinetic and elastic energy densities
 \beq{EacslDef}
 \Eacsl = \frac12 \rhosl \omega^2 \avr{u_k u_k} + \frac12 \avr{\epsilon_{ik}\sigma_{ik}},
 \eeq
with summation over the repeated indices $i,k = x,y,z$, and where $\epsilon_{ik} = \frac12(\pp_i u_k+\pp_k u_i)$ is the strain tensor components, while  $\avr{AB} = \frac12 \re[A^* B]$ is the time average over one oscillation period of the fields $A$ and $B$ given in the complex time-harmonic notation.

The fluid (water) of density $\rhofl$ and sound speed $\cfl$, with an acoustic pressure $p$ and acoustic velocity $\vvv$ at angular frequency $\omega$, is modeled as pressure acoustics with a weak absorption $\Gamma_\mr{fl} \ll 1$, but no viscocity, \cite{Ley2017}
 \bsub
 \eqlab{gov_eq_fluid}
 \bal
 \nabla^2 p &= -\frac{\omega^2}{\cfl^2} (1+\mathrm{i}\Gamma_\mr{fl})^2\: p,\\
 \vvv &= \frac{-\ii}{\omega\rhofl}\:\nablabf p.
 \eal
 \esub
The acoustic energy density $\Eacfl$ in the fluid domain is the sum of the kinetic and compressional energy densities,
 \beq{EacflDef}
 \Eacfl = \frac12 \rhofl \avr{v_k v_k} + \frac12 \kapfl \avr{p^2},
 \eeq
where $\kapfl = (\rhofl\cflsqr)^{-1}$ is the compressibility of the fluid.

\subsection{Boundary conditions and fluid-solid coupling}

The applied boundary conditions are the usual ones \cite{Ley2017}, namely that (1) the stress and the velocity fields are continuous across all fluid-solid interfaces, which then provides the coupling between the fluid and solid domains, (2) the stress is zero on all outer boundaries facing the air, and (3) the piezoelectric actuation is represented by a given displacement in the normal direction at that part of the solid surface, where the actuator is attached. The influence ($\leftarrow$) from the surroundings on a given domain with an outward-pointing surface normal $\nnn$ is
 \bsub
 \eqlab{bcAll}
 \begin{alignat}{2}
 \eqlab{bcFLfromSL}
 \text{Fluid domain} & \leftarrow  \text{solid:}
 & \vvv \cdot \nnn  &=  -\ii \omega\: \uuu\cdot\nnn \\
 \eqlab{bcSLfromFL}
 \text{Solid domain} & \leftarrow  \text{fluid:}
 & \sigmabfsl \cdot \nnn &= - p\: \nnn,   \\ %
 \eqlab{bcSLfromPZT}
 \text{Solid domain} & \leftarrow  \text{transducer: }
 & \uuu &= \pm d_0\:\nnn,\\
 \eqlab{bcSLfromAIR}
 \text{Solid domain} & \leftarrow  \text{air:}
 & \sigmabfsl \cdot \nnn &= \zerovec,
 \end{alignat}
 \esub

\subsection{The acoustic radiation force on suspended microparticles}

In a good acoustofluidic device, the acoustic radiation force $\FFFrad$ on a suspended microparticle should be sufficiently large. In this work we consider 10-$\SImum$-diameter spherical polystyrene "Styron 666" (ps) particles with density $\rhoPS$ and compressibility $\kapPS$. For such a large microparticle suspended in water of density $\rhofl$ and compressibility $\kapfl$, thermoviscous boundary layers can be neglected, and the monopole and dipole acoustic scattering coefficients $f_0$ and $f_1$ are given by \cite{Karlsen2015},
 \beq{f0f1def}
 f_0 = 1 - \frac{\kapPS}{\kapfl} = 0.468, \qquad f_1 = \frac{2(\rhoPS-\rhofl)}{2\rhoPS + \rhofl} = 0.034.
 \eeq
In the presence of an acoustic field~\eqnoref{gov_eq_fluid} of pressure $p$ and velocity $\vvv$, the suspended microparticle experience an acoustic radiation force $\FFFrad$ given by~\cite{Settnes2012}
 \bal
 \eqlab{Frad_Settnes}
 \FFFrad \!= - \pi a^3\! \left(\dfrac{2}{3}\:\kapfl
 \re\!\left[f_0^* p^* \nablabf p \right] - \rhofl\!
 \re\!\left[f_1^* \vec{v}^* \!\cdot\! \nablabf \vec{v} \right] \right).
 \eal
Here, $a$ is the particle radius, and the asterisk denotes complex conjugation.

\subsection{Numerical implementation}
\seclab{Numerics}
Following the procedure described in Ref.~\cite{Ley2017}, including mesh convergence tests, the coupled field equations~\eqsnoref{gov_eq_solid}{gov_eq_fluid} for the fluid pressure $p$ and the elastic-solid displacement $\uuu$, subject to the boundary conditions \eqref{bcAll}, are implemented and solved on weak form using the finite-element solver Comsol Multiphysics 5.3a \cite{Comsol53a}. Lastly, to evaluate the quality of the acoustophoresis in the given device, the last step is to use \eqsref{f0f1def}{Frad_Settnes} to compute the acoustic scattering coefficients $f_0$ and $f_1$ as well as the acoustic radiation force $\FFFrad$  acting on a single 10-$\SImum$-diameter spherical polystyrene tracer bead suspended at different positions in the water-filled microchannel.

\section{Results for the 2D model of a conventional hard-walled device}
\seclab{Results2D_hard}

We begin our analysis with the computational less demanding and faster simulations restricted to 2D cross sections of the systems, before moving on to the more heavy, full 3D simulations. Moreover, to establish a baseline for evaluating the results for the acoustophoretic capabilities of all-polymer devices, we first analyze an ideal hard-walled device. Then we progress to a conventional, acoustically hard, but elastic silicon-glass device, before finally treating an all-polymer device.

\subsection{An ideal hard-walled device in 2D}
\seclab{hard_wall_2D_device}

We consider a rectangular channel cross section of width $\Wfl$ and height $\Hfl$ in the vertical $y$-$z$ plane and centered around $y=0$. If its walls are infinitely hard, it supports a standing half-wave pressure resonance $p^{{}}_\mr{hard}$ in the horizontal $y$-direction at the frequency $f^{{}}_\mr{hard}$. The resonance is characterized by its amplitude $p^{{}}_\mr{ac}$ and corresponding acoustic energy density $E^{{}}_\mr{hard}$, and it results in a radiation force $\FFFrad_\mr{hard}$ of amplitude $\Frad_\mr{ac}$ on a given spherical tracer particle of radius $a$ and acoustic monopole and dipole scattering coefficients $f_0$ and $f_1$ suspended in the channel. Using the dimensions of the fluid domain listed in \tabref{geometry_Si_Py}, the resonance properties can be summarized as,~\cite{Barnkob2012a}
 \bsub
 \eqlab{HardResonance}
 \bal
 \eqlab{fhard}
 f^{{}}_\mr{hard} &= \frac{\cfl}{2\Wfl} = 1.985~\SIMHz,\\
 \eqlab{phard}
 p^{{}}_\mr{hard} &= p^{{}}_\mr{ac}\sin\big(k_y y\big),\: \text{ with } k_y = \frac{\pi}{\Wfl},\\
 \eqlab{Ehard}
 E^{{}}_\mr{hard} &= \frac14 \kapfl p^{{2}}_\mr{ac},\\
 \eqlab{Fradhard}
 \FFFrad_\mr{hard} &= -\Frad_\mr{ac}\sin\big(2k_y y\big)\: \een_y,\\
 \nn &
 \quad \text{with } \Frad_\mr{ac} = \bigg[\frac{f_0}{3} + \frac{f_1}{2}\bigg]4\pi^2\frac{a^3E^{{}}_\mr{hard}}{\Wfl}.
 \eal
 \esub
This hard-walled resonance is completely decoupled from the motion of the surrounding solid. Its standing pressure wave $p^{{}}_\mr{hard}$ is a perfect sinusoidal half-wave with a vertical nodal line at the channel center, and the radiation force $\FFFrad_\mr{hard}$ is similarly a perfect sinusoidal full-wave that pushes suspending particles horizontally towards their stable equilibrium positions at the vertical center plane of the channel. No vertical force is exerted.

The $f^{{}}_\mr{hard}$ resonance is an idealization of the standing pressure half wave used in many acoustophoresis experiments. In the following study, we choose this resonance as the prime example of an acoustic resonance that leads to particularly good acoustophoresis.

\begin{table}[t]
\centering
\caption{\tablab{geometry_Si_Py} The length scales of the rectangular silicon-Pyrex (si-py) system with a fluid-filled rectangular channel (fl). The values are from Ref.~\cite{Barnkob2010}, except $\Delta W$ introduced in \figref{DeviceGeom}(b).}
\begin{ruledtabular}
\begin{tabular}{lcc}
Dimension & Solid domain & Fluid domain \\ \hline
Length & $\Lsl = 50~\SImm$    & $\Lfl = 40~\SImm$  \rule{0mm}{1.1em} \\
Width  & $\Wsl = 2.52~\SImm$  & $\Wfl = 377~\SImum$ \\
Height & $\Hsi = 350~\SImum$  & $\Hfl = 157~\SImum$ \\
Height & $\Hpy = 1130~\SImum$ & ---                 \\
Actuator gap & $\Delta W = 100~\SImum$ & ---
\end{tabular}
\end{ruledtabular}
\end{table}

\subsection{A conventional silicon-glass device in 2D}
\seclab{SiPy_2D_device}

As mentioned in the introduction, silicon-glass devices have successfully been applied to microscale acoustofluidic tasks \cite{Thevoz2010, Augustsson2012, Zmijan2015, Ohlin2015, Augustsson2016, Ohlsson2016}. Following the experiments of Refs.~\cite{Barnkob2010, Augustsson2011, Barnkob2012a, Muller2013}, we choose here to study the long, straight rectangular channel of length $\Lfl$, width $\Wfl$, and height $\Hfl$ fabricated by KOH etch into the surface of a rectangular $\avr{100}$ silicon (Si) wafer with length $\Lsl$, width $\Wsl$, and  height $\Hsi = \Hslbot + \Hfl$, which is sealed off with a borosilicate glass lid (Pyrex) of height $\Hpy = \Hsltop$ and the same length $\Lsl$ and width $\Wsl$. In the following, we restrict our modeling to the 2D vertical cross section in the $y$-$z$ plane sketched in \figref{DeviceGeom}(a) using the device dimensions listed in \tabref{geometry_Si_Py} and the anti-symmetric actuation of amplitude $d_0 = 0.1$~nm on the bottom surface as sketched in \figref{DeviceGeom}(b) and given in \eqref{bcSLfromPZT}.

Most of the theoretical analysis of the acoustofluidic properties of this microchannel has been carried out using the assumption of infinitely hard walls \cite{Muller2012, Muller2014, Muller2015}. This assumption is justified for silicon-glass devices, as these solids are heavier and stiffer than water, as quantified by the ratio $\Zsl/\Zfl$ of the specific acoustic impedances listed in \tabsref{material_param_sl}{material_param_fl},
  \beq{Zratios}
  \frac{Z^\mr{py}_\mr{sl}}{Z^\mr{wa}_\mr{fl}} =  8.35, \qquad
  \frac{Z^\mr{si}_\mr{sl}}{Z^\mr{wa}_\mr{fl}} = 13.16.
  \eeq
However, as in Ref.~\cite{Bach2018}, we now include the elastic properties of the surrounding elastic silicon and Pyrex material using the parameter values listed in \tabref{material_param_sl}.

\begin{figure}[t]
\centering
\includegraphics[width=\columnwidth]{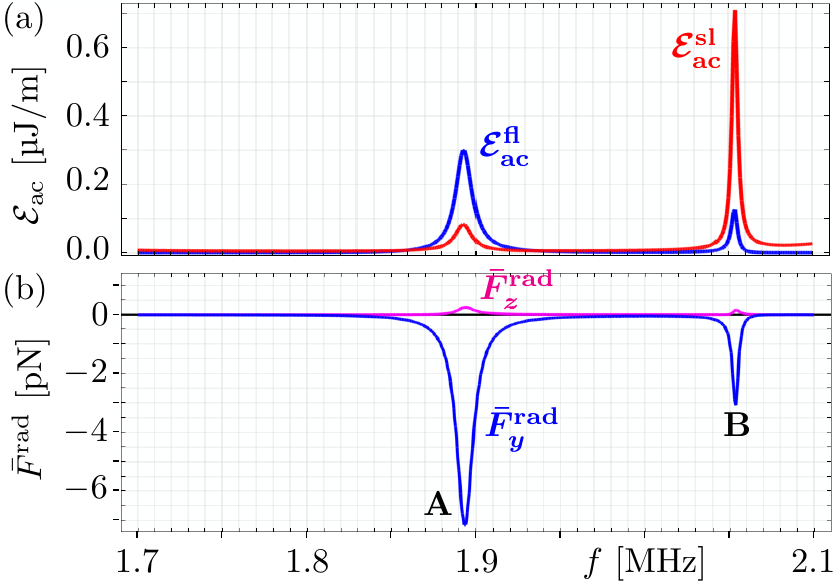}
\caption{\figlab{Si_Py_spectra} Numerical results for the frequency dependency of the rectangular silicon-glass device actuated at frequency $f$ using the anti-symmetric actuation of \figref{DeviceGeom}(b) with displacement amplitude $d_0 = 0.1$~nm. Two strong resonances A and B are identified. (a) Line plot of the total acoustic energy $\calEacfl$ (blue) in the fluid domain $\Omegafl$ and $\calEacsl$ (red) in the solid domain $\Omegasl$  versus $f$ for the silicon-glass device. (b) The spatially averaged radiation force components $\Fradbar_y$ (blue) and $\Fradbar_z$ (magenta).}
\end{figure}

According to \eqref{fhard}, the hard-walled device has a well-defined half-wave resonance at $f^{{}}_\mr{hard} = 1.985~\SIMHz$. We therefore simulate the silicon-glass device in the frequency range from 1.7 to 2.1~MHz, see \figref{Si_Py_spectra}. We compute the frequency dependency of the acoustic energy $\calEacsl$ and $\calEacfl$ in the solid and fluid domain, defined by
 \beq{calEacSlFl}
 \calEacsl = \int_\Omegasl \Eacsl\: \dm y \dm z, \qquad
 \calEacfl = \int_\Omegafl \Eacfl\: \dm y \dm z.
 \eeq
as well as of the spatial average $\FFFradbar$ of the acoustic radiation force on a 10-$\SImum$-diameter test particle in the fluid domain,
 \bsub
 \eqlab{FradbarDef}
 \bal
 \eqlab{FradbarYdef}
 \Fradbar_y &= \frac{1}{\Wfl\Hfl}\int_\Omegafl \frac{y}{|y|}\:\Frad_y\: \dm y \dm z,\\
 \eqlab{FradbarZdef}
 \Fradbar_z &= \frac{1}{\Wfl\Hfl}\int_\Omegafl \big|\Frad_z\big|\: \dm y \dm z.
 \eal
 \esub
Here, we have "rectified" the $y$-component average through the anti-symmetric pre-factor $y/|y|$ to obtain a large value of $\Fradbar_y$, when $\Frad_y$ has the useful anti-symmetric form, and where we have taken the absolute value of $\Frad_z$ before averaging, so that a minimal value of $\Fradbar_z$ is obtained, when the magnitude of the vertical component of the acoustic radiation force is small everywhere. This is done to help identifying resonances having  a behavior similar to the conventional and very useful one of the hard-walled device, see \eqref{HardResonance}.

\begin{table}[t]
\centering
\caption{\tablab{Si_Py_resonances} The fluid-domain-averaged acoustic energy density $\Eacfl$, the components of the average acoustic radiation force $\FFFradbar$ and the figure of merit $R$ for the resonances A and B of \figref{Si_Py_spectra} in the silicon-glass device of width $\Wsl = 2.52$~mm and height $\Hsl = 1.48$~mm, and $d_0 = 0.1$~nm.}
\begin{ruledtabular}
\begin{tabular}{ccrrrr}
Resonance & Frequency & $\Eacfl$ & $\Fradbar_y$ & $\Fradbar_z$ & $R\;\;$ \\
number & [MHz] & [Pa]  & [pN] & [pN] & [1]$\;$  \\ \hline
A & 1.893 & 5.07 & 7.15 & 0.25 & \rule{0mm}{1.1em} 29.1\\
B & 2.054 & 2.15 & 3.08 & 0.17 &  18.1
\end{tabular}
\end{ruledtabular}
\end{table}

\begin{figure*}[t]
\centering
\includegraphics[]{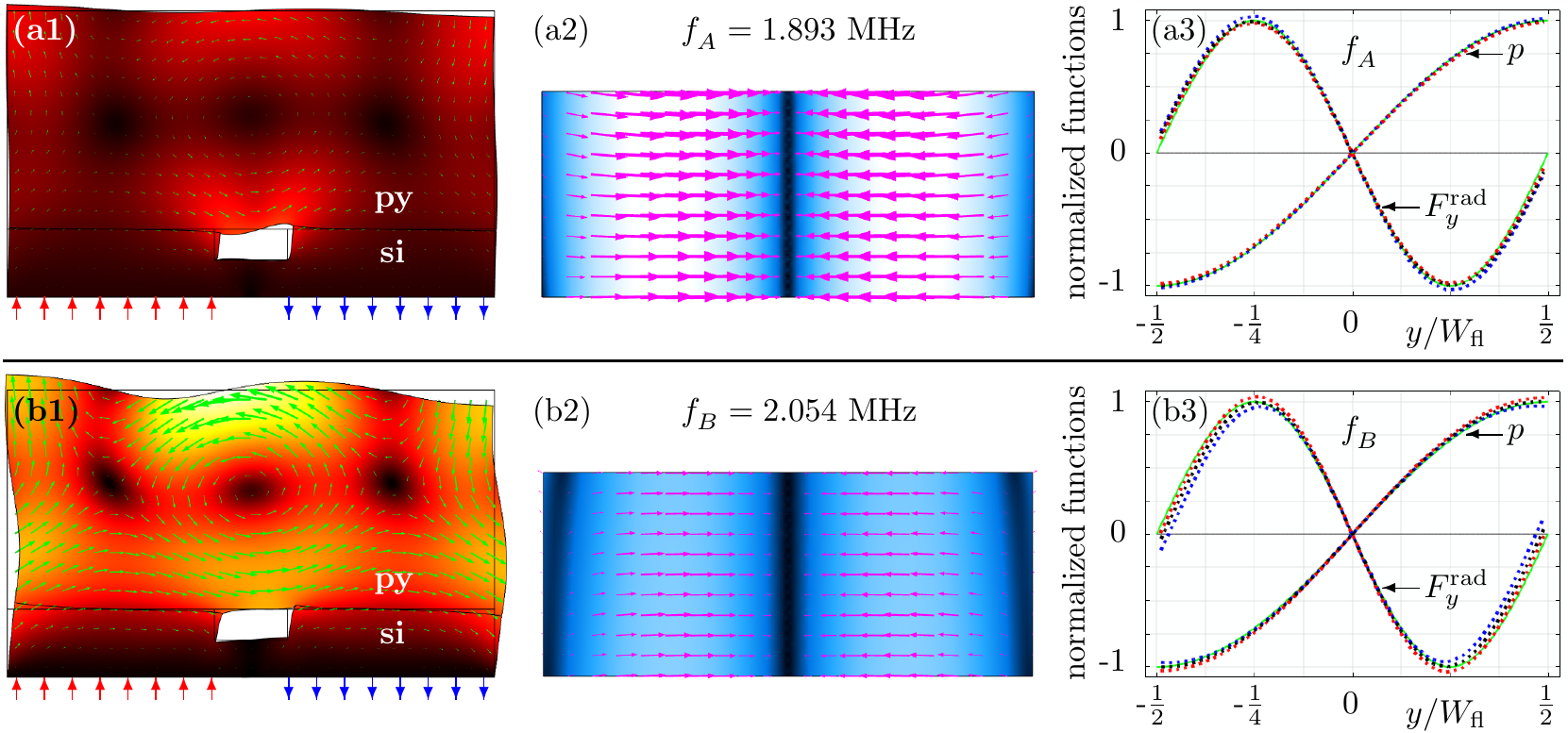}
\caption{\figlab{SiPy_res_A_B}
Numerical results for resonance A and B in \figref{Si_Py_spectra} of the glass-silicon device. Top row (a1), (a2), and (a3) are for $\fA = 1.893$~MHz. (a1) Vector plot in the solid domain $\Omegasl$  of the displacement field $\uuu$ (green arrows) and color plot of its amplitude $u$ from 0~nm (black) to 2.6~nm (white). To be visible, the nm-scale displacement has been increased by a factor 50,000.  (a2) The radiation force $\FFFrad$ on a 10-$\SImum$-diameter spherical polystyrene tracer particle as a function of its position in the fluid domain $\Omegafl$. Color plot of the magnitude $\Frad$ from 0~pN (black) to 11.5~pN (white) and vector plot (magenta arrows) of $\FFFrad$. (a3) Comparison of the numerical simulation of the silicon-glass device (dotted lines) of \figref{Si_Py_spectra} with the analytical result of the hard-wall device (solid green line). The normalized pressure $p/\pnorm$ (half wave, $\pnorm = 209$~kPa) and the $y$-component  $\Frad_y/\Fradnorm$  of the radiation force (full wave, $\Fradnorm = 11.0$~pN) plotted along the top, center, and bottom horizontal lines at $z=0.9\Hfl$ (blue dots), $z=0.5\Hfl$ (black dots), and $z=0.1\Hfl$ (red dots), respectively. Bottom row (b1), (b2), and (b3) are as top row (a1), (a2), and (a3), respectively, but for $\fB = 2.054$~MHz with $\pnorm = 138$~kPa and $\Fradnorm = 5.05$~pN.}
\end{figure*}

The results for $\calEacsl$, $\calEacsl$, $\Fradbar_y$, and $\Fradbar_z$ versus frequency are shown in \figref{Si_Py_spectra}. We readily identify two strong resonances A and B that show up in all four quantities at $\fA = 1.893$~MHz and $\fB = 2.054$~MHz, respectively. Acoustically, these resonances are different. In \figref{Si_Py_spectra}(a) we see that for resonance~A, the acoustic energy $\calEacfl = 0.300$~J/m in the fluid is 3.7 times larger than $\calEacsl = 0.082$~J/m in the solid, in spite of the area $\calA_\mr{fl} = 0.059~\SImm^2$ of the fluid domain is 62 times smaller than the area  $\calA_\mr{sl}  = 3.670~\SImm^2$ of the solid domain. Thus, A may be characterized as a fluid-domain resonance. Conversely, for resonance~B, $\calEacfl = 0.127$~J/m in the fluid is 0.18 times $\calEacsl = 0.710$~J/m in the solid, and B may thus be characterized as a whole-system resonance.  In \figref{Si_Py_spectra}(b) we observe that for both resonance A and B, the magnitude $\Fradbar_y$ of the horizontal component of $\FFFradbar$ is much larger than that of the vertical component $\Fradbar_z$. To quantify this, we introduce the figure of merit $R$, listed in \tabref{Si_Py_resonances}, as the ratio
 \beq{meritRdef}
 R = -\frac{\Fradbar_y}{\Fradbar_z}.
 \eeq
$R$ will be large in situations, where the acoustic radiation force has the desired property of a large anti-symmetric horizontal component and small vertical component. For the two resonances the figure of merit is $R^{{}}_A = 29.1$ and $R^{{}}_B  = 18.1$, respectively, indicating that the fluid-domain resonance A has better acoustophoretic properties than the whole-system resonance B. We also note that the predicted acoustic energy density $\Eac$ at the two resonances listed in \tabref{Si_Py_resonances}, resulting from the chosen actuation amplitude $d_0 = 0.1$~nm, are in agreement with typical experimental values that fall in the range from 1 to 100~Pa \cite{Barnkob2010, Augustsson2011, Barnkob2012a, Muller2013}.

We study resonance A and B in more detail in \figref{SiPy_res_A_B}. The color and vector plots in Figs.~\fignoref{SiPy_res_A_B}(a1) and \fignoref{SiPy_res_A_B}(b1) reveal that of the displacement $\uuu$ for resonance~A is much larger at the interface to the channel than at the outer surface, while for resonance~B the opposite is true. The color and vector plot of the acoustic radiation force $\FFFrad$ for resonance~A in \figref{SiPy_res_A_B}(a2), shows that it is nearly identical to the perfectly horizontal radiation force $\FFFrad_\mr{hard}$ in \eqref{Fradhard} of the hard-walled device with its vertical line of stability in the center of the channel. As seen in \figref{SiPy_res_A_B}(b2), resonance~B is weaker, but is quite similar except for some skewness at the sides. Finally, through the horizontal line plots of the radiation force $\Frad_y$ and the pressure $p$ at the top, center, and bottom heights $z=0.9\Hfl$, $0.5\Hfl$, and $0.1\Hfl$ shown in Figs.~\fignoref{SiPy_res_A_B}(a3) and~\fignoref{SiPy_res_A_B}(b3), we compare more closely resonance $\fA$ and $\fB$ in the silicon-glass device with the ideal half-wave resonance $f^{{}}_\mr{hard}$ in the hard-walled device. It is seen how closely the pressure $p$ and the $y$-component $\Frad_y$ of the radiation force at the resonance $\fA$  reproduce the corresponding analytical expressions \eqsref{phard}{Fradhard} of the resonance $f^{{}}_\mr{hard}$. Resonance $\fB$  deviates a little more from resonance $f^{{}}_\mr{hard}$.

In conclusion, \figsref{Si_Py_spectra}{SiPy_res_A_B} demonstrate than in an acoustically hard silicon-glass device, two types of resonances may exist. One type is the fluid-domain resonance, represented by resonance~A, defined by a high acoustic energy in the fluid domain and relatively little coupling to the solid domain. Compared to the actuation displacement $d_0$, the amplitude $u$ of displacement field is relatively large at the inner fluid-solid interface and small at the outer surface of the solid. The resulting acoustic fields and radiation force closely resemble those of the ideal hard-walled case. Conversely, the other type is the whole-system resonance, represented by resonance~B, defined by a high acoustic energy in the solid domain and a low one in the fluid domain. The displacement at the outer surface of the solid is larger than at the inner fluid-solid interface. The resulting acoustic fields and radiation force deviate a little from those of the ideal hard-walled case.

\section{Results for the 2D model of an all-polymer PMMA device}
\seclab{Results2D_PMMA}

We now apply our model to the study of an acoustically soft PMMA device. We do expect to see a different behavior compared to the acoustically hard silicon-glass device, because of the small ratio $\Zsl/\Zfl$ of the specific acoustic impedances listed in \tabsref{material_param_sl}{material_param_fl},
  \beq{ZratioPMMA}
  \frac{Z^\mr{PMMA}_\mr{sl}}{Z^\mr{wa}_\mr{fl}} =  2.14,
  \eeq
which is 4 to 7 times smaller than those of pyrex and silicon given in \eqref{Zratios}. In particular, we expect the fluid-domain resonances of type A to vanish in this case, leaving only whole-system ultrasound resonances (WSUR) of type B.

\subsection{Analysis of the 3-mm wide PMMA device}

To investigate this hypothesis, we perform numerical simulations on a PMMA device with the dimensions listed in \tabref{geometry_PMMA}. These values refer to the preliminary experimental work carried out by Pelle Ohlsson and Ola Jakobsson at the company AcouSort AB in Lund, Sweden, on acoustophoresis in all-polymer devices \cite{AcouSort}. The device is fabricated from a thin bottom-layer PMMA film of height $\Hslbot$ and width $\Wsl$, see \figref{DeviceGeom}(a), onto which is bonded a thick PMMA block of  height $\Hsltop + \Hfl$ and same width $\Wsl$ that contains a rectangular channel of height $\Hfl$ and width $\Wfl$ embossed or milled into its bottom surface. In the model we again use the anti-symmetric actuation of \figref{DeviceGeom}(b), typically in the range from 1 to 2.1~MHz, but now with the larger actuation displacement amplitude $d_0 = 0.3$~nm to mimic the softer material.

\begin{table}[b]
\centering
\caption{\tablab{geometry_PMMA} The length scales of the rectangular PMMA system (sl) with a fluid-filled rectangular channel (fl). The values are provided by AcouSort~\cite{AcouSort}.}
\begin{ruledtabular}
\begin{tabular}{lcc}
Dimension & Solid domain & Fluid domain \\ \hline
Length & $\Lsl = 50~\SImm$             & $\Lfl = 40~\SImm$  \rule{0mm}{1.1em} \\
Width  & $\Wsl = 3.0~\SImm$            & $\Wfl = 375~\SImum$ \\
Height & $\Hslbot = 175~\SImum$        & $\Hfl = 150~\SImum$ \\
Height & $\Hsltop+\Hfl = 1000~\SImum$  & ---                 \\
Actuator gap & $\Delta W = 100~\SImum$ & ---
\end{tabular}
\end{ruledtabular}
\end{table}

The results for $\calEacsl$, $\calEacsl$, $\Fradbar_y$, and $\Fradbar_z$ versus frequency in the PMMA device are shown in \figref{PMMA_spectra}. Right away, we notice one striking difference between these spectra and the one for the silicon-glass device in \figref{Si_Py_spectra}. All the observed resonances are whole-system resonances of type B. The energy $\calEacsl$ of the solid is one to two orders of magnitude larger than the energy $\calEacfl$ of the fluid domain. In fact, in \figref{PMMA_spectra}(a) we need to use a logarithmic scale to be able to see these two energies in the same plot. In \figref{PMMA_spectra}(b) we plot the average acoustic radiation force components $\Fradbar_y$ and $\Fradbar_z$, and as before observe that a number of resonances are clearly identified in all four quantities.

\begin{figure}[t]
\centering
\includegraphics[width=\columnwidth]{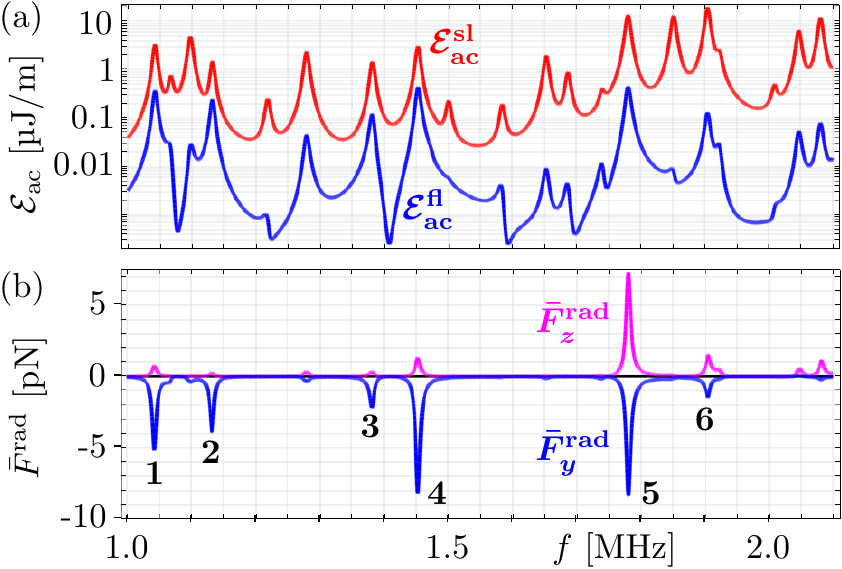}
\caption{\figlab{PMMA_spectra}  Numerical results for the rectangular PMMA device actuated at frequency $f$ using the anti-symmetric actuation of \figref{DeviceGeom}(b) with displacement amplitude $d_0 = 0.3$~nm. (a) Log-lin plot of the total acoustic energy $\calEacfl$ (blue) in the fluid domain $\Omegafl$ and $\calEacsl$ (red) in the solid domain $\Omegasl$ versus actuation frequency $f$. (b) The spatially averaged radiation force components $\Fradbar_y$ (blue line) and $\Fradbar_z$ (magenta line) as a function of the actuation frequency $f$ showing six prominent resonances 1--6.}
\end{figure}

\begin{table}[b]
\centering
\caption{\tablab{PMMA_resonances} The fluid-domain-averaged acoustic energy density $\Eacfl$, the components of the average acoustic radiation force $\FFFradbar$ and the figure of merit $R$ for each of the six pronounced resonances of \figref{PMMA_spectra} in the 2D PMMA device of width $\Wsl = 3$~mm and height $\Hsl = 1.175$~mm, and $d_0 = 0.3$~nm.}
\begin{ruledtabular}
\begin{tabular}{ccrrrr}
Resonance & Frequency & $\Eacfl$ & $\Fradbar_y$ & $\Fradbar_z$ & $R\;\;$ \\
number & [MHz] & [Pa]  & [pN] & [pN] & [1]$\;$  \\ \hline
1 & 1.0425 & 6.63 & 5.19 & 0.78 & \rule{0mm}{1.1em} 6.7\\
2 & 1.1320 & 4.40 & 3.91 & 0.26 & 14.9\\
3 & 1.3815 & 2.14 & 2.21 & 0.34 &  6.6\\
4 & 1.4530 & 7.72 & 8.22 & 1.34 &  6.1\\
5 & 1.7810 & 7.93 & 8.36 & 7.27 &  1.2\\
6 & 1.9045 & 2.32 & 1.50 & 1.48 &  1.0
\end{tabular}
\end{ruledtabular}
\end{table}

The six most prominent resonances and their figure of merit $R$ are listed in \tabref{PMMA_resonances}. Based on these data, we predict that resonance~2 at $f_2 = 1.1320$~MHz with the highest figure of merit, $R_2 = 14.9$, has properties resembling those of the ideal resonance $f^{{}}_\mr{hard}$ the most, while resonance~6 at $f_6 = 1.9045$~MHz, the one closets to the ideal resonance frequency $f_\mr{hard} = 2.0$~MHz, is not good given its low figure of merits $R_6 = 1.0$.

\begin{figure*}[t]
\centering
\includegraphics[width=\textwidth]{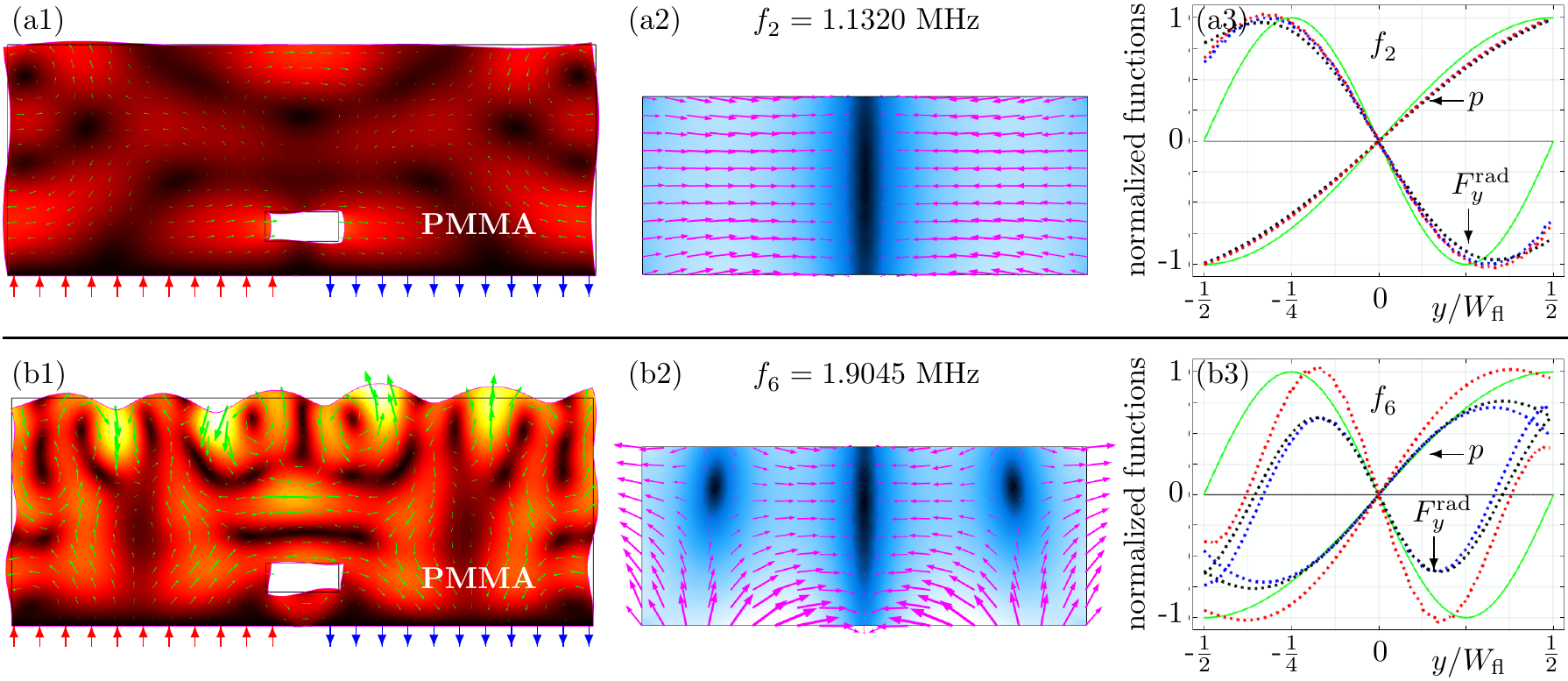}
\caption{\figlab{PMMA_res_2_6}  Numerical results for resonance 2 and 6 in \figref{PMMA_spectra} of the PMMA device. Top row (a1), (a2), and (a3) are for $f_2 = 1.1320$~MHz. (a1) Vector plot in the solid domain $\Omegasl$  of the displacement field $\uuu$ (green arrows) and color plot of its amplitude $u$ from 0~nm (black) to 25~nm (light yellow). To be visible, the nm-scale displacement has been increased by a factor 3000.  (a2) The radiation force $\FFFrad$ on a 10-$\SImum$-diameter spherical polystyrene tracer particle as a function of its position in the fluid domain $\Omegafl$. Color plot of the magnitude $\Frad$ from 0~pN (black) to 9~pN (white) and vector plot (magenta arrows) of $\FFFrad$. (a3) The normalized pressure $p/\pnorm$ (half wave, $\pnorm = 150$~kPa) and the $y$-component  $\Frad_y/\Fradnorm$  of the radiation force (full wave, $\Fradnorm = 5.4$~pN) along the top, center, and bottom horizontal lines at $z=0.9\Hfl$ (blue dots), $z=0.5\Hfl$ (black dots), and $z=0.1\Hfl$ (red dots), respectively. Bottom row (b1), (b2), and (b3) are as top row (a1), (a2), and (a3), respectively, but for $f_6 = 1.9045$~MHz with $\pnorm = 135$~kPa and $\Fradnorm = 6.8$~pN.}
\end{figure*}

This hypothesis is checked in \figref{PMMA_res_2_6}. We see that indeed most of the physical properties of PMMA resonance~2 in \figref{PMMA_res_2_6}(a1), (a2), and (a3) are similar to the almost perfect silicon-glass resonance $\fA$. In one aspect it is even superior: The radiation force in resonance $f_2$ is not zero at the channel walls $y = \pm\frac12 \Wfl$ as is the case with $\fA$ and $f^{{}}_\mr{hard}$. In this sense $f_2$ is better to move particles from any starting point in the channel towards the vertical nodal line at $y=0$. In contrast, the plots in \figref{PMMA_res_2_6}(b1), (b2), and~(b3) show that the acoustics in the channel, induced by the whole-system resonance in the PMMA device, results in a radiation force that only in the middle part of the fluid domain points towards the center line. At the edges it points the opposite way. Also a strong vertical component is observed. Clearly in this case, the low figure of merit is correctly predicting a resonance not well suited for acoustophoretic applications corresponding to the acoustically hard system.

With this example, we have demonstrated the main point of the paper: Microchannels embedded in acoustically sort materials are not able to support a resonance close to the ideal hard-wall resonance. However, whole-system resonances, primarily defined by having a relatively big displacement field in the large solid domain, may nevertheless induce a pressure field in the small fluid domain, which have properties suitable for acoustophoretic applications. We have identified these few acoustophoretically useful whole-system resonances as those peaks in the spectra of the energy and the radiation force components that have the largest figure of merit $R$.

\subsection{The width dependence of the PMMA device}

We further demonstrate the use the above whole-system-ultrasound-resonance principle, by studying the width dependency of the resonances in the PMMA device. In \figref{Frad_vs_Ws_f0} we show a scatter plot of resonances as a function of the device width $\Wsl$ from 1 to 6~mm and of the actuation frequency $f_0$ from 1 to 2.1~MHz. The area of each data point is proportional to the magnitude $\Fradbar_y$. The maximal amplitude is $\Fradbar_y = 337$~pN obtained for $\Wsl = 1.3$~mm and $f = 1.328$~MHz, and we have left out resonances with a magnitude lower than $\Fradbar_y = 0.3$~pN. The color code of the points goes from black at $\Fradbar = 0.03$~pN to yellow at $\Fradbar = 30$~pN.

\begin{figure}[t]
\centering
\includegraphics[width=0.95\columnwidth]{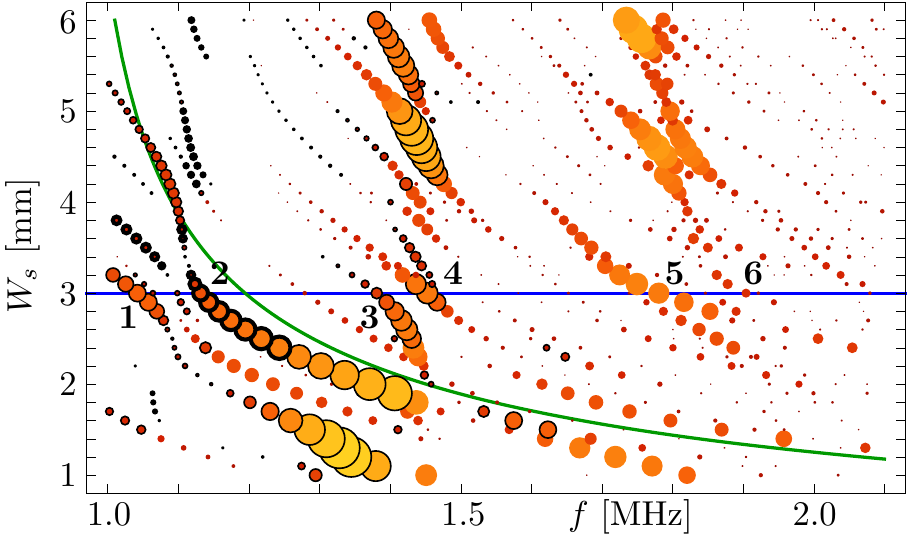}
\caption{\figlab{Frad_vs_Ws_f0} The average radiation force component $\Fradbar_y$ in the fluid domain in a PMMA device at resonance  for a 10-$\SImum$-diameter polystyrene tracer particle plotted as a colored scatter plot from 0.03~pN (black) to 30~pN (yellow) versus the polymer chip width $\Wsl$ and the actuation frequency $f$. The the area of each data point is proportional to the magnitude of $\Fradbar_y$. Resonances with a figure of merit $5 \leq R < 10$ and $10 \leq R$ are further marked with a thin black and thick black rim, respectively. The horizontal blue line and the resonances $1, 2, \ldots, 6$ correspond to the line plot in \figref{PMMA_spectra}. The green curve is the simplified analytical prediction $W^{(\frac12,1)}_\mr{sl}$, \eqref{fHalfOne}.}
\end{figure}

To visualize the figure of merit $R$ of the resonance points in \figref{Frad_vs_Ws_f0}, we have added a thick black rim to the resonance points with $R\geq 10$ and a thin black rim to points with  $5 \leq R <10$. The highest figure of merit is $R = 27.3$ obtained for $\Wsl = 3.5$~mm and $f = 1.054$~MHz. Using such a plot, we are able to efficiently map out the parameter space in the hunt for suitable device designs for making all-polymer devices able to produce good acoustophoresis.

It is clear from \figref{Frad_vs_Ws_f0} that the acoustophoretically good resonance~2 lies on a curve of resonance points for which the resonance frequency increases as the device width $\Wsl$ decreases. To understand this behavior, we study more closely the displacement field of resonance~2 in \figref{PMMA_res_2_6}(a1). We note that not only is there half a shear wave along the $y$ axis due to the actuation, but there is also a full shear wave along the $z$ axis. This resoance we denote the $(\frac12$,1)-resonance. In a simplistic model of shear-wave propagation of velocity $\cT$ decoupled from the longitudinal wave, we predict the resonance frequency $f_{(\frac12,1})$ for the $(\frac12,1)$-resonance to be
 \beq{fHalfOne}
 f_{(\frac12,1)} = \cT \sqrt{\frac{2^2}{W^2_\mr{sl}} + \frac{1^2}{H^2_\mr{sl}}} \quad \text{or} \quad
 W^{(\frac12,1)}_\mr{sl}(f) = \frac{2}{\sqrt{\frac{f^2}{\cTsqr} - \frac{1}{H^2_\mr{sl}}}}.
 \eeq
As seen from \figref{Frad_vs_Ws_f0} (green curve), this na\"ive prediction is in fair agreement with the line of resonant points to which resonance~2 belongs. At the position of the fluid domain, this particular shear wave results in a horizontal oscillating displacement field which is compatible with ideal standing pressure half-wave in the channel.

\section{Results for the 3D model of an all-polymer PMMA device}
\seclab{Results3D}

A more realistic test of the whole-system-resonance principle for a device in 3D is shown in \figref{Results3D} for a microchannel of length $\Lfl = 40$~mm embedded in PMMA chip of length $\Lsl = 50$~mm and with the same 2D cross section as above, see \tabref{geometry_PMMA}.  The ultrasound actuation of amplitude $d_0 = 0.3$~nm is anti-symmetric in the $y$ direction and independent of $x$ along the entire bottom as sketched \figref{DeviceGeom}(b). The whole-system-resonance principle is now used to identify resonances useful for acoustophoresis in this coupled polymer-water systems.

\begin{widetext}
\mbox{}\noindent
\begin{figure}[h!]
\centering
\includegraphics[width=\textwidth]{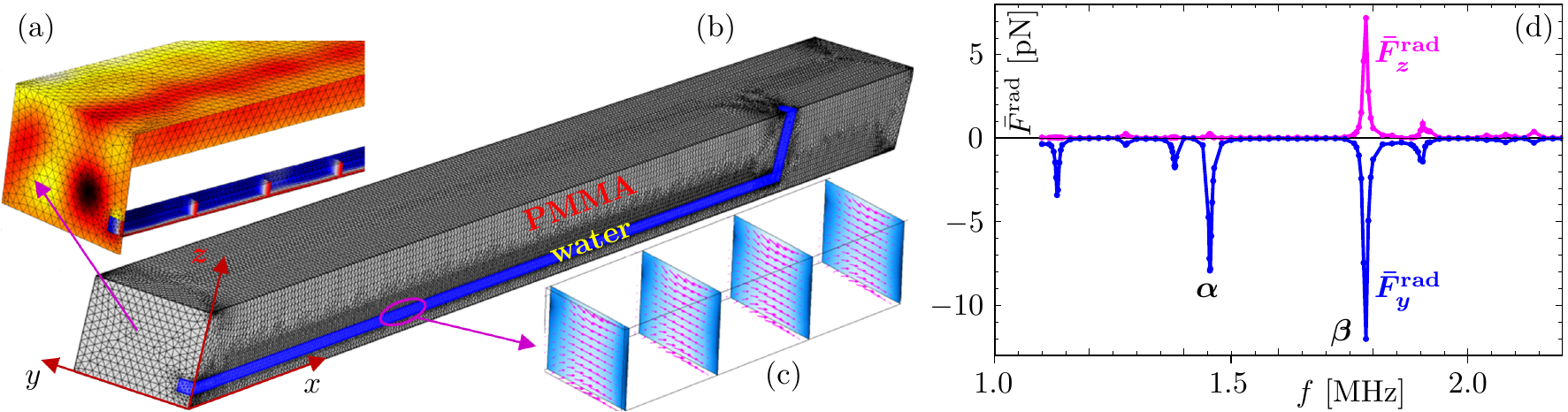}
\caption{\figlab{Results3D} Numerical simulation of a water-filled microchannel of length $\Lfl = 40$~mm, width $\Wfl = 375~\SImum$, and height $\Hfl = 150~\SImum$ embedded in a PMMA chip of length $\Lsl = 50$~mm, width $\Wsl = 3~\SImm$, and height $\Hsl = 1175~\SImum$, and actuated at the resonance frequency $f = 1.455$~MHz. Due to the symmetry at the $yz$- and $xz$-plane, only the quarter domain $0<x<\frac12\Lsl$ and $0<y<\frac12\Wsl$ is simulated as in Ref.~\cite{Ley2017}. The anti-symmetric actuation of \figref{DeviceGeom}(b) is used with amplitude $d_0 = 0.3$~nm. (a) Color plots of the amplitude of the displacement $u$ from 0~nm (black) to 4.5~nm (white) on the outer PMMA surface and the pressure pressure $p$ from 0~kPa red to 190~kPa (blue) near the symmetry plane $x=0$ of the system. (b) The finite-element mesh (thin gray lines) used in the simulation. (c) The radiation force $\FFFrad$ on 10-$\SImum$-diameter polystyrene tracer particles in a short section of the microchannel. (d) The frequency dependency of the volume-averaged acoustic radiation force components $\Fradbar_y$ and $\Fradbar_z$ showing two pronounced resonances $\alpha$ and $\beta$ at $f_\alpha = 1.455$~MHz and $f_\beta = 1.785$~MHz.}
\end{figure}
\end{widetext}

\noindent
The frequency dependent volume-averaged acoustic radiation force $\Fradbar_y$ is shown in \figref{Results3D}(d), and it reveals two strong resonances $\alpha$ and $\beta$ at $f_\alpha = 1.455$~MHz and at $f_\beta = 1.785$~MHz with properties listed in \tabref{PMMA_resonances_3D}. Their respective figures of merit $R_\alpha =  32.5$ and $R_\beta = 1.7$ indicate that resonance $\alpha$ is more likely to have good properties for acoustophoresis. This is verified by the detailed structure of the pressure field in \figref{Results3D}(a) and the acoustic radiation force $\FFFrad$ represented by four vertical cut-planes in \figref{Results3D}(c). The acoustic energy density at resonance $\alpha$ is predicted to be $\Eacfl = 4.0$~Pa for the assumed actuation amplitude $d_0 = 0.3$~nm.

\begin{table}[t]
\centering
\caption{\tablab{PMMA_resonances_3D} The fluid-domain-averaged acoustic energy density $\Eacfl$, the components of the average acoustic radiation force $\FFFradbar$ and the figure of merit $R$ for resonances $\alpha$ and $\beta$ \figref{Results3D}(d) in the 3D PMMA device of length $\Lsl = 50$~mm, width $\Wsl = 3$~mm and height $\Hsl = 1.175$~mm, and anti-symmetric actuation of amplitude $d_0 = 0.3$~nm.}
\begin{ruledtabular}
\begin{tabular}{ccrrrr}
Resonance & Frequency & $\Eacfl$ & $\Fradbar_y$ & $\Fradbar_z$ & $R\;\;$ \\
number & [MHz] & [Pa]  & [pN] & [pN] & [1]$\;$  \\ \hline
$\alpha$ & 1.455 & 4.00 & 7.94 & 0.24 & \rule{0mm}{1.1em} 32.5\\
$\beta$  & 1.785 & 3.63 & 12.06 & 7.18 &  1.7
\end{tabular}
\end{ruledtabular}
\end{table}

Preliminary experiments performed by Pelle Ohlson and Ola Jakobsson at AcouSort AB in Lund, Sweden, on PMMA devices nominally identical to the one simulated here, have confirmed the existence of a whole-system ultrasound resonance at $f = 1.55$~MHz with an acoustic energy density of $\Eacfl = 12$~Pa, fully capable of obtaining acoustophoretic focusing on suspended 10-$\SImum$-diameter polystyrene tracer particles \cite{AcouSort}. The predicted whole-system resonance frequency $f_\alpha = 1.455$~MHz is only 6~\% lower than the observed one, but 27~\% lower than the hard-wall resonance, and the predicted acoustic energy density $\Eac = 4$~Pa will equal the observed one, if the assumed actuation amplitude is increased by a factor of $\sqrt{12~\SIPa / 4~\SIPa} = 1.7$ from $d_0 = 0.3$~nm to $d_0 = 0.5$~nm. A detailed report of these experiments and their comparison with simulations will be given elsewhere.

\section{Concluding discussion}
\seclab{Conclusion}

We have presented a numerical study of an acoustically soft device consisting of a fluid channel inside a PMMA chip, where a single split or two separate ultrasound transducers operate in anti-phase to excite a standing ultrasound wave in the entire polymer chip. The model takes into account the fully coupled longitudinal and transverse displacement waves in the solid domain and their coupling with the pressure field in the fluid domain.

The nearly identical specific acoustic impedances for PMMA (and other polymers) and water does not allow for localized resonances in the water domain that are decoupled from the solid domain, as is usually the case in acoustically hard systems such as the conventional silicon-glass devices. Thus for all-polymer systems, the conventional thinking in terms of standing half-wave resonances in the fluid domain cannot be maintained. Instead, the acoustic fields in the two domains are strongly coupled, and given its large volume compared to the fluid domain, the resonance behavior is mainly determined by the solid domain.

Nevertheless, we have found whole-system ultrasound resonances (WSUR) that supports good acoustophoretic action in the water channel. These WSUR resonances can be identified theoretically by computing resonance peaks in the area-averaged (in 2D) or volume-averaged (in 3D) acoustophoretic force component $\Fradbar_y$ by \eqref{FradbarDef} in combination with the figure of merit $R$ defined in \eqref{meritRdef}. We have demonstrated in both 2D and 3D models, how such WSUR resonance indeed have acoustic properties that are comparable to those found in conventional systems used for acoustophoresis.

Preliminary experiments performed at AcouSort AB \cite{AcouSort} on the specific PMMA system modeled here in 3D, have verified the existence of WSUR resonance and their ability to generate good acoustophoresis of a quality fully comparable to that obtained in conventional silicon-glass devices.

In the analysis presented here, we have focused on acoustophoresis similar to the one obtained by the simple standing pressure half-wave resonance that focuses suspended particles in the vertical center plane. However, our method is not restricted to this particular type of resonances. It is straightforward to extend it to searching for whole-system resonances with other spatial structures simply by changing the figure of merit to one that reflects the wanted type of resonance. We believe that the whole-system-ultrasound-resonance principle presented in this paper has the potential of becoming an important design tool in the development of high-quality, all-polymer, acoustofluidic devices.

\section*{Acknowledgements}
RM was supported by the People Programme (Marie Curie Actions) of the European Union's Seventh Framework Programme (FP7/2007-2013) under REA grant agreement no.~609405 (COFUNDPostdocDTU). We are grateful for the discussions on experimental realizations of our ideas that we have had with Pelle Ohlsson, Ola Jakobsson, and Torsten Freltoft from AcouSort AB, and with Thomas Laurell at Lund University.


%
%


\begin{thebibliography}{63}%
\makeatletter
\providecommand \@ifxundefined [1]{%
 \@ifx{#1\undefined}
}%
\providecommand \@ifnum [1]{%
 \ifnum #1\expandafter \@firstoftwo
 \else \expandafter \@secondoftwo
 \fi
}%
\providecommand \@ifx [1]{%
 \ifx #1\expandafter \@firstoftwo
 \else \expandafter \@secondoftwo
 \fi
}%
\providecommand \natexlab [1]{#1}%
\providecommand \enquote  [1]{``#1''}%
\providecommand \bibnamefont  [1]{#1}%
\providecommand \bibfnamefont [1]{#1}%
\providecommand \citenamefont [1]{#1}%
\providecommand \href@noop [0]{\@secondoftwo}%
\providecommand \href [0]{\begingroup \@sanitize@url \@href}%
\providecommand \@href[1]{\@@startlink{#1}\@@href}%
\providecommand \@@href[1]{\endgroup#1\@@endlink}%
\providecommand \@sanitize@url [0]{\catcode `\\12\catcode `\$12\catcode
  `\&12\catcode `\#12\catcode `\^12\catcode `\_12\catcode `\%12\relax}%
\providecommand \@@startlink[1]{}%
\providecommand \@@endlink[0]{}%
\providecommand \url  [0]{\begingroup\@sanitize@url \@url }%
\providecommand \@url [1]{\endgroup\@href {#1}{\urlprefix }}%
\providecommand \urlprefix  [0]{URL }%
\providecommand \Eprint [0]{\href }%
\providecommand \doibase [0]{http://dx.doi.org/}%
\providecommand \selectlanguage [0]{\@gobble}%
\providecommand \bibinfo  [0]{\@secondoftwo}%
\providecommand \bibfield  [0]{\@secondoftwo}%
\providecommand \translation [1]{[#1]}%
\providecommand \BibitemOpen [0]{}%
\providecommand \bibitemStop [0]{}%
\providecommand \bibitemNoStop [0]{.\EOS\space}%
\providecommand \EOS [0]{\spacefactor3000\relax}%
\providecommand \BibitemShut  [1]{\csname bibitem#1\endcsname}%
\let\auto@bib@innerbib\@empty
\bibitem [{\citenamefont {Lenshof}\ \emph {et~al.}(2012)\citenamefont
  {Lenshof}, \citenamefont {Magnusson},\ and\ \citenamefont
  {Laurell}}]{Lenshof2012a}%
  \BibitemOpen
  \bibfield  {author} {\bibinfo {author} {\bibfnamefont {A.}~\bibnamefont
  {Lenshof}}, \bibinfo {author} {\bibfnamefont {C.}~\bibnamefont {Magnusson}},
  \ and\ \bibinfo {author} {\bibfnamefont {T.}~\bibnamefont {Laurell}},\
  }\bibfield  {title} {\bibinfo {title} {Acoustofluidics 8: Applications in
  acoustophoresis in continuous flow microsystems},\ }\href {\doibase
  10.1039/c2lc21256k} {\bibfield  {journal} {\bibinfo  {journal} {Lab Chip}\
  }\textbf {\bibinfo {volume} {12}},\ \bibinfo {pages} {1210} (\bibinfo {year}
  {2012})}\BibitemShut {NoStop}%
\bibitem [{\citenamefont {Gedge}\ and\ \citenamefont {Hill}(2012)}]{Gedge2012}%
  \BibitemOpen
  \bibfield  {author} {\bibinfo {author} {\bibfnamefont {M.}~\bibnamefont
  {Gedge}}\ and\ \bibinfo {author} {\bibfnamefont {M.}~\bibnamefont {Hill}},\
  }\bibfield  {title} {\bibinfo {title} {Acoustofluidics 17: Surface acoustic
  wave devices for particle manipulation},\ }\href {\doibase
  10.1039/C2LC40565B} {\bibfield  {journal} {\bibinfo  {journal} {Lab Chip}\
  }\textbf {\bibinfo {volume} {12}},\ \bibinfo {pages} {2998} (\bibinfo {year}
  {2012})}\BibitemShut {NoStop}%
\bibitem [{\citenamefont {Sackmann}\ \emph {et~al.}(2014)\citenamefont
  {Sackmann}, \citenamefont {Fulton},\ and\ \citenamefont
  {Beebe}}]{Sackmann2014}%
  \BibitemOpen
  \bibfield  {author} {\bibinfo {author} {\bibfnamefont {E.~K.}\ \bibnamefont
  {Sackmann}}, \bibinfo {author} {\bibfnamefont {A.~L.}\ \bibnamefont
  {Fulton}}, \ and\ \bibinfo {author} {\bibfnamefont {D.~J.}\ \bibnamefont
  {Beebe}},\ }\bibfield  {title} {\bibinfo {title} {The present and future role
  of microfluidics in biomedical research},\ }\href {\doibase
  10.1038/nature13118} {\bibfield  {journal} {\bibinfo  {journal} {Nature}\
  }\textbf {\bibinfo {volume} {507}},\ \bibinfo {pages} {181} (\bibinfo {year}
  {2014})}\BibitemShut {NoStop}%
\bibitem [{\citenamefont {Laurell}\ and\ \citenamefont
  {Lenshof}(2015)}]{Laurell2014}%
  \BibitemOpen
  \bibinfo {editor} {\bibfnamefont {T.}~\bibnamefont {Laurell}}\ and\ \bibinfo
  {editor} {\bibfnamefont {A.}~\bibnamefont {Lenshof}},\ eds.,\ \href@noop {}
  {\emph {\bibinfo {title} {{Microscale Acoustofluidics}}}}\ (\bibinfo
  {publisher} {Royal Society of Chemistry},\ \bibinfo {address} {Cambridge},\
  \bibinfo {year} {2015})\BibitemShut {NoStop}%
\bibitem [{\citenamefont {Antfolk}\ and\ \citenamefont
  {Laurell}(2017)}]{Antfolk2017}%
  \BibitemOpen
  \bibfield  {author} {\bibinfo {author} {\bibfnamefont {M.}~\bibnamefont
  {Antfolk}}\ and\ \bibinfo {author} {\bibfnamefont {T.}~\bibnamefont
  {Laurell}},\ }\bibfield  {title} {\bibinfo {title} {Continuous flow
  microfluidic separation and processing of rare cells and bioparticles found
  in blood - {A} review},\ }\href {\doibase 10.1016/j.aca.2017.02.017}
  {\bibfield  {journal} {\bibinfo  {journal} {Anal. Chim. Acta}\ }\textbf
  {\bibinfo {volume} {965}},\ \bibinfo {pages} {9} (\bibinfo {year}
  {2017})}\BibitemShut {NoStop}%
\bibitem [{\citenamefont {Thevoz}\ \emph {et~al.}(2010)\citenamefont {Thevoz},
  \citenamefont {Adams}, \citenamefont {Shea}, \citenamefont {Bruus},\ and\
  \citenamefont {Soh}}]{Thevoz2010}%
  \BibitemOpen
  \bibfield  {author} {\bibinfo {author} {\bibfnamefont {P.}~\bibnamefont
  {Thevoz}}, \bibinfo {author} {\bibfnamefont {J.~D.}\ \bibnamefont {Adams}},
  \bibinfo {author} {\bibfnamefont {H.}~\bibnamefont {Shea}}, \bibinfo {author}
  {\bibfnamefont {H.}~\bibnamefont {Bruus}}, \ and\ \bibinfo {author}
  {\bibfnamefont {H.~T.}\ \bibnamefont {Soh}},\ }\bibfield  {title} {\bibinfo
  {title} {Acoustophoretic synchronization of mammalian cells in
  microchannels},\ }\href {\doibase 10.1021/ac100357u} {\bibfield  {journal}
  {\bibinfo  {journal} {Anal. Chem.}\ }\textbf {\bibinfo {volume} {82}},\
  \bibinfo {pages} {3094} (\bibinfo {year} {2010})}\BibitemShut {NoStop}%
\bibitem [{\citenamefont {Augustsson}\ \emph {et~al.}(2012)\citenamefont
  {Augustsson}, \citenamefont {Magnusson}, \citenamefont {Nordin},
  \citenamefont {Lilja},\ and\ \citenamefont {Laurell}}]{Augustsson2012}%
  \BibitemOpen
  \bibfield  {author} {\bibinfo {author} {\bibfnamefont {P.}~\bibnamefont
  {Augustsson}}, \bibinfo {author} {\bibfnamefont {C.}~\bibnamefont
  {Magnusson}}, \bibinfo {author} {\bibfnamefont {M.}~\bibnamefont {Nordin}},
  \bibinfo {author} {\bibfnamefont {H.}~\bibnamefont {Lilja}}, \ and\ \bibinfo
  {author} {\bibfnamefont {T.}~\bibnamefont {Laurell}},\ }\bibfield  {title}
  {\bibinfo {title} {Microfluidic, label-free enrichment of prostate cancer
  cells in blood based on acoustophoresis},\ }\href {\doibase
  10.1021/ac301723s} {\bibfield  {journal} {\bibinfo  {journal} {Anal. Chem.}\
  }\textbf {\bibinfo {volume} {84}},\ \bibinfo {pages} {7954} (\bibinfo {year}
  {2012})}\BibitemShut {NoStop}%
\bibitem [{\citenamefont {Zmijan}\ \emph {et~al.}(2015)\citenamefont {Zmijan},
  \citenamefont {Jonnalagadda}, \citenamefont {Carugo}, \citenamefont {Kochi},
  \citenamefont {Lemm}, \citenamefont {Packham}, \citenamefont {Hill},\ and\
  \citenamefont {Glynne-Jones}}]{Zmijan2015}%
  \BibitemOpen
  \bibfield  {author} {\bibinfo {author} {\bibfnamefont {R.}~\bibnamefont
  {Zmijan}}, \bibinfo {author} {\bibfnamefont {U.~S.}\ \bibnamefont
  {Jonnalagadda}}, \bibinfo {author} {\bibfnamefont {D.}~\bibnamefont
  {Carugo}}, \bibinfo {author} {\bibfnamefont {Y.}~\bibnamefont {Kochi}},
  \bibinfo {author} {\bibfnamefont {E.}~\bibnamefont {Lemm}}, \bibinfo {author}
  {\bibfnamefont {G.}~\bibnamefont {Packham}}, \bibinfo {author} {\bibfnamefont
  {M.}~\bibnamefont {Hill}}, \ and\ \bibinfo {author} {\bibfnamefont
  {P.}~\bibnamefont {Glynne-Jones}},\ }\bibfield  {title} {\bibinfo {title}
  {High throughput imaging cytometer with acoustic focussing},\ }\href
  {\doibase 10.1039/c5ra19497k} {\bibfield  {journal} {\bibinfo  {journal}
  {{RSC} Advances}\ }\textbf {\bibinfo {volume} {5}},\ \bibinfo {pages} {83206}
  (\bibinfo {year} {2015})}\BibitemShut {NoStop}%
\bibitem [{\citenamefont {Ohlin}\ \emph {et~al.}(2015)\citenamefont {Ohlin},
  \citenamefont {Iranmanesh}, \citenamefont {Christakou},\ and\ \citenamefont
  {Wiklund}}]{Ohlin2015}%
  \BibitemOpen
  \bibfield  {author} {\bibinfo {author} {\bibfnamefont {M.}~\bibnamefont
  {Ohlin}}, \bibinfo {author} {\bibfnamefont {I.}~\bibnamefont {Iranmanesh}},
  \bibinfo {author} {\bibfnamefont {A.~E.}\ \bibnamefont {Christakou}}, \ and\
  \bibinfo {author} {\bibfnamefont {M.}~\bibnamefont {Wiklund}},\ }\bibfield
  {title} {\bibinfo {title} {Temperature-controlled mpa-pressure ultrasonic
  cell manipulation in a microfluidic chip},\ }\href {\doibase
  10.1039/c5lc00490j} {\bibfield  {journal} {\bibinfo  {journal} {Lab Chip}\
  }\textbf {\bibinfo {volume} {15}},\ \bibinfo {pages} {3341} (\bibinfo {year}
  {2015})}\BibitemShut {NoStop}%
\bibitem [{\citenamefont {Collins}\ \emph {et~al.}(2015)\citenamefont
  {Collins}, \citenamefont {Morahan}, \citenamefont {Garcia-Bustos},
  \citenamefont {Doerig}, \citenamefont {Plebanski},\ and\ \citenamefont
  {Neild}}]{Collins2015}%
  \BibitemOpen
  \bibfield  {author} {\bibinfo {author} {\bibfnamefont {D.~J.}\ \bibnamefont
  {Collins}}, \bibinfo {author} {\bibfnamefont {B.}~\bibnamefont {Morahan}},
  \bibinfo {author} {\bibfnamefont {J.}~\bibnamefont {Garcia-Bustos}}, \bibinfo
  {author} {\bibfnamefont {C.}~\bibnamefont {Doerig}}, \bibinfo {author}
  {\bibfnamefont {M.}~\bibnamefont {Plebanski}}, \ and\ \bibinfo {author}
  {\bibfnamefont {A.}~\bibnamefont {Neild}},\ }\bibfield  {title} {\bibinfo
  {title} {Two-dimensional single-cell patterning with one cell per well driven
  by surface acoustic waves},\ }\href {\doibase 10.1038/ncomms9686} {\bibfield
  {journal} {\bibinfo  {journal} {Nat. Commun.}\ }\textbf {\bibinfo {volume}
  {6}},\ \bibinfo {pages} {8686} (\bibinfo {year} {2015})}\BibitemShut
  {NoStop}%
\bibitem [{\citenamefont {Guo}\ \emph {et~al.}(2016)\citenamefont {Guo},
  \citenamefont {Mao}, \citenamefont {Chen}, \citenamefont {Xie}, \citenamefont
  {Lata}, \citenamefont {Li}, \citenamefont {Ren}, \citenamefont {Liu},
  \citenamefont {Yang}, \citenamefont {Dao}, \citenamefont {Suresh},\ and\
  \citenamefont {Huang}}]{Guo2016}%
  \BibitemOpen
  \bibfield  {author} {\bibinfo {author} {\bibfnamefont {F.}~\bibnamefont
  {Guo}}, \bibinfo {author} {\bibfnamefont {Z.}~\bibnamefont {Mao}}, \bibinfo
  {author} {\bibfnamefont {Y.}~\bibnamefont {Chen}}, \bibinfo {author}
  {\bibfnamefont {Z.}~\bibnamefont {Xie}}, \bibinfo {author} {\bibfnamefont
  {J.~P.}\ \bibnamefont {Lata}}, \bibinfo {author} {\bibfnamefont
  {P.}~\bibnamefont {Li}}, \bibinfo {author} {\bibfnamefont {L.}~\bibnamefont
  {Ren}}, \bibinfo {author} {\bibfnamefont {J.}~\bibnamefont {Liu}}, \bibinfo
  {author} {\bibfnamefont {J.}~\bibnamefont {Yang}}, \bibinfo {author}
  {\bibfnamefont {M.}~\bibnamefont {Dao}}, \bibinfo {author} {\bibfnamefont
  {S.}~\bibnamefont {Suresh}}, \ and\ \bibinfo {author} {\bibfnamefont {T.~J.}\
  \bibnamefont {Huang}},\ }\bibfield  {title} {\bibinfo {title}
  {Three-dimensional manipulation of single cells using surface acoustic
  waves},\ }\href {\doibase 10.1073/pnas.1524813113} {\bibfield  {journal}
  {\bibinfo  {journal} {PNAS}\ }\textbf {\bibinfo {volume} {113}},\ \bibinfo
  {pages} {1522} (\bibinfo {year} {2016})}\BibitemShut {NoStop}%
\bibitem [{\citenamefont {Augustsson}\ \emph {et~al.}(2016)\citenamefont
  {Augustsson}, \citenamefont {Karlsen}, \citenamefont {Su}, \citenamefont
  {Bruus},\ and\ \citenamefont {Voldman}}]{Augustsson2016}%
  \BibitemOpen
  \bibfield  {author} {\bibinfo {author} {\bibfnamefont {P.}~\bibnamefont
  {Augustsson}}, \bibinfo {author} {\bibfnamefont {J.~T.}\ \bibnamefont
  {Karlsen}}, \bibinfo {author} {\bibfnamefont {H.-W.}\ \bibnamefont {Su}},
  \bibinfo {author} {\bibfnamefont {H.}~\bibnamefont {Bruus}}, \ and\ \bibinfo
  {author} {\bibfnamefont {J.}~\bibnamefont {Voldman}},\ }\bibfield  {title}
  {\bibinfo {title} {Iso-acoustic focusing of cells for size-insensitive
  acousto-mechanical phenotyping},\ }\href {\doibase 10.1038/ncomms11556}
  {\bibfield  {journal} {\bibinfo  {journal} {Nat. Commun.}\ }\textbf {\bibinfo
  {volume} {7}},\ \bibinfo {pages} {11556} (\bibinfo {year}
  {2016})}\BibitemShut {NoStop}%
\bibitem [{\citenamefont {Ohlsson}\ \emph {et~al.}(2016)\citenamefont
  {Ohlsson}, \citenamefont {Evander}, \citenamefont {Petersson}, \citenamefont
  {Mellhammar}, \citenamefont {Lehmusvuori}, \citenamefont {Karhunen},
  \citenamefont {Soikkeli}, \citenamefont {Seppa}, \citenamefont {Tuunainen},
  \citenamefont {Spangar}, \citenamefont {von Lode}, \citenamefont
  {Rantakokko-Jalava}, \citenamefont {Otto}, \citenamefont {Scheding},
  \citenamefont {Soukka}, \citenamefont {Wittfooth},\ and\ \citenamefont
  {Laurell}}]{Ohlsson2016}%
  \BibitemOpen
  \bibfield  {author} {\bibinfo {author} {\bibfnamefont {P.}~\bibnamefont
  {Ohlsson}}, \bibinfo {author} {\bibfnamefont {M.}~\bibnamefont {Evander}},
  \bibinfo {author} {\bibfnamefont {K.}~\bibnamefont {Petersson}}, \bibinfo
  {author} {\bibfnamefont {L.}~\bibnamefont {Mellhammar}}, \bibinfo {author}
  {\bibfnamefont {A.}~\bibnamefont {Lehmusvuori}}, \bibinfo {author}
  {\bibfnamefont {U.}~\bibnamefont {Karhunen}}, \bibinfo {author}
  {\bibfnamefont {M.}~\bibnamefont {Soikkeli}}, \bibinfo {author}
  {\bibfnamefont {T.}~\bibnamefont {Seppa}}, \bibinfo {author} {\bibfnamefont
  {E.}~\bibnamefont {Tuunainen}}, \bibinfo {author} {\bibfnamefont
  {A.}~\bibnamefont {Spangar}}, \bibinfo {author} {\bibfnamefont
  {P.}~\bibnamefont {von Lode}}, \bibinfo {author} {\bibfnamefont
  {K.}~\bibnamefont {Rantakokko-Jalava}}, \bibinfo {author} {\bibfnamefont
  {G.}~\bibnamefont {Otto}}, \bibinfo {author} {\bibfnamefont {S.}~\bibnamefont
  {Scheding}}, \bibinfo {author} {\bibfnamefont {T.}~\bibnamefont {Soukka}},
  \bibinfo {author} {\bibfnamefont {S.}~\bibnamefont {Wittfooth}}, \ and\
  \bibinfo {author} {\bibfnamefont {T.}~\bibnamefont {Laurell}},\ }\bibfield
  {title} {\bibinfo {title} {Integrated acoustic separation, enrichment, and
  microchip polymerase chain reaction detection of bacteria from blood for
  rapid sepsis diagnostics},\ }\href {\doibase 10.1021/acs.analchem.6b00323}
  {\bibfield  {journal} {\bibinfo  {journal} {Anal. Chem.}\ }\textbf {\bibinfo
  {volume} {88}},\ \bibinfo {pages} {9403} (\bibinfo {year}
  {2016})}\BibitemShut {NoStop}%
\bibitem [{\citenamefont {Hammarstr\"{o}m}\ \emph {et~al.}(2014)\citenamefont
  {Hammarstr\"{o}m}, \citenamefont {Nilson}, \citenamefont {Laurell},
  \citenamefont {Nilsson},\ and\ \citenamefont
  {Ekstr\"{o}m}}]{Hammarstrom2014a}%
  \BibitemOpen
  \bibfield  {author} {\bibinfo {author} {\bibfnamefont {B.}~\bibnamefont
  {Hammarstr\"{o}m}}, \bibinfo {author} {\bibfnamefont {B.}~\bibnamefont
  {Nilson}}, \bibinfo {author} {\bibfnamefont {T.}~\bibnamefont {Laurell}},
  \bibinfo {author} {\bibfnamefont {J.}~\bibnamefont {Nilsson}}, \ and\
  \bibinfo {author} {\bibfnamefont {S.}~\bibnamefont {Ekstr\"{o}m}},\
  }\bibfield  {title} {\bibinfo {title} {Acoustic trapping for bacteria
  identification in positive blood cultures with maldi-tof ms},\ }\href
  {\doibase 10.1021/ac502020f} {\bibfield  {journal} {\bibinfo  {journal}
  {Anal. Chem.}\ }\textbf {\bibinfo {volume} {86}},\ \bibinfo {pages} {10560}
  (\bibinfo {year} {2014})}\BibitemShut {NoStop}%
\bibitem [{\citenamefont {Sitters}\ \emph {et~al.}(2015)\citenamefont
  {Sitters}, \citenamefont {Kamsma}, \citenamefont {Thalhammer}, \citenamefont
  {Ritsch-Marte}, \citenamefont {Peterman},\ and\ \citenamefont
  {Wuite}}]{Sitters2015}%
  \BibitemOpen
  \bibfield  {author} {\bibinfo {author} {\bibfnamefont {G.}~\bibnamefont
  {Sitters}}, \bibinfo {author} {\bibfnamefont {D.}~\bibnamefont {Kamsma}},
  \bibinfo {author} {\bibfnamefont {G.}~\bibnamefont {Thalhammer}}, \bibinfo
  {author} {\bibfnamefont {M.}~\bibnamefont {Ritsch-Marte}}, \bibinfo {author}
  {\bibfnamefont {E.~J.~G.}\ \bibnamefont {Peterman}}, \ and\ \bibinfo {author}
  {\bibfnamefont {G.~J.~L.}\ \bibnamefont {Wuite}},\ }\bibfield  {title}
  {\bibinfo {title} {Acoustic force spectroscopy},\ }\href {\doibase
  10.1038/nmeth.3183} {\bibfield  {journal} {\bibinfo  {journal} {Nat. Meth.}\
  }\textbf {\bibinfo {volume} {12}},\ \bibinfo {pages} {47} (\bibinfo {year}
  {2015})}\BibitemShut {NoStop}%
\bibitem [{\citenamefont {Drinkwater}(2016)}]{Drinkwater2016}%
  \BibitemOpen
  \bibfield  {author} {\bibinfo {author} {\bibfnamefont {B.~W.}\ \bibnamefont
  {Drinkwater}},\ }\bibfield  {title} {\bibinfo {title} {Dynamic-field devices
  for the ultrasonic manipulation of microparticles},\ }\href {\doibase
  10.1039/C6LC00502K} {\bibfield  {journal} {\bibinfo  {journal} {Lab Chip}\
  }\textbf {\bibinfo {volume} {16}},\ \bibinfo {pages} {2360} (\bibinfo {year}
  {2016})}\BibitemShut {NoStop}%
\bibitem [{\citenamefont {Collins}\ \emph {et~al.}(2016)\citenamefont
  {Collins}, \citenamefont {Devendran}, \citenamefont {Ma}, \citenamefont {Ng},
  \citenamefont {Neild},\ and\ \citenamefont {Ai}}]{Collins2016}%
  \BibitemOpen
  \bibfield  {author} {\bibinfo {author} {\bibfnamefont {D.~J.}\ \bibnamefont
  {Collins}}, \bibinfo {author} {\bibfnamefont {C.}~\bibnamefont {Devendran}},
  \bibinfo {author} {\bibfnamefont {Z.}~\bibnamefont {Ma}}, \bibinfo {author}
  {\bibfnamefont {J.~W.}\ \bibnamefont {Ng}}, \bibinfo {author} {\bibfnamefont
  {A.}~\bibnamefont {Neild}}, \ and\ \bibinfo {author} {\bibfnamefont
  {Y.}~\bibnamefont {Ai}},\ }\bibfield  {title} {\bibinfo {title} {Acoustic
  tweezers via sub-time-of-flight regime surface acoustic waves},\ }\href
  {\doibase 10.1126/sciadv.1600089} {\bibfield  {journal} {\bibinfo  {journal}
  {Science Advances}\ }\textbf {\bibinfo {volume} {2}},\ \bibinfo {pages}
  {e1600089} (\bibinfo {year} {2016})}\BibitemShut {NoStop}%
\bibitem [{\citenamefont {Lim}\ \emph {et~al.}(2016)\citenamefont {Lim},
  \citenamefont {Li}, \citenamefont {Lin}, \citenamefont {Yoon}, \citenamefont
  {Lee}, \citenamefont {Jung}, \citenamefont {Chow},\ and\ \citenamefont
  {Shung}}]{Lim2016}%
  \BibitemOpen
  \bibfield  {author} {\bibinfo {author} {\bibfnamefont {H.~G.}\ \bibnamefont
  {Lim}}, \bibinfo {author} {\bibfnamefont {Y.}~\bibnamefont {Li}}, \bibinfo
  {author} {\bibfnamefont {M.-Y.}\ \bibnamefont {Lin}}, \bibinfo {author}
  {\bibfnamefont {C.}~\bibnamefont {Yoon}}, \bibinfo {author} {\bibfnamefont
  {C.}~\bibnamefont {Lee}}, \bibinfo {author} {\bibfnamefont {H.}~\bibnamefont
  {Jung}}, \bibinfo {author} {\bibfnamefont {R.~H.}\ \bibnamefont {Chow}}, \
  and\ \bibinfo {author} {\bibfnamefont {K.~K.}\ \bibnamefont {Shung}},\
  }\bibfield  {title} {\bibinfo {title} {Calibration of trapping force on
  cell-size objects from ultrahigh-frequency single-beam acoustic tweezer},\
  }\href {\doibase 10.1109/TUFFC.2016.2600748} {\bibfield  {journal} {\bibinfo
  {journal} {{IEEE T. Ultrason. Ferr.}}\ }\textbf {\bibinfo {volume} {63}},\
  \bibinfo {pages} {1988} (\bibinfo {year} {2016})}\BibitemShut {NoStop}%
\bibitem [{\citenamefont {Baresch}\ \emph {et~al.}(2016)\citenamefont
  {Baresch}, \citenamefont {Thomas},\ and\ \citenamefont
  {Marchiano}}]{Baresch2016}%
  \BibitemOpen
  \bibfield  {author} {\bibinfo {author} {\bibfnamefont {D.}~\bibnamefont
  {Baresch}}, \bibinfo {author} {\bibfnamefont {J.-L.}\ \bibnamefont {Thomas}},
  \ and\ \bibinfo {author} {\bibfnamefont {R.}~\bibnamefont {Marchiano}},\
  }\bibfield  {title} {\bibinfo {title} {Observation of a single-beam gradient
  force acoustical trap for elastic particles: Acoustical tweezers},\ }\href
  {\doibase 10.1103/PhysRevLett.116.024301} {\bibfield  {journal} {\bibinfo
  {journal} {Phys. Rev. Lett.}\ }\textbf {\bibinfo {volume} {116}},\ \bibinfo
  {pages} {024301} (\bibinfo {year} {2016})}\BibitemShut {NoStop}%
\bibitem [{\citenamefont {Gautam}\ \emph {et~al.}(2018)\citenamefont {Gautam},
  \citenamefont {Burger}, \citenamefont {Wilcox}, \citenamefont {Cumbo},
  \citenamefont {Graves},\ and\ \citenamefont {Piyasena}}]{Gautam2018}%
  \BibitemOpen
  \bibfield  {author} {\bibinfo {author} {\bibfnamefont {G.~P.}\ \bibnamefont
  {Gautam}}, \bibinfo {author} {\bibfnamefont {T.}~\bibnamefont {Burger}},
  \bibinfo {author} {\bibfnamefont {A.}~\bibnamefont {Wilcox}}, \bibinfo
  {author} {\bibfnamefont {M.~J.}\ \bibnamefont {Cumbo}}, \bibinfo {author}
  {\bibfnamefont {S.~W.}\ \bibnamefont {Graves}}, \ and\ \bibinfo {author}
  {\bibfnamefont {M.~E.}\ \bibnamefont {Piyasena}},\ }\bibfield  {title}
  {\bibinfo {title} {Simple and inexpensive micromachined aluminum microfluidic
  devices for acoustic focusing of particles and cells},\ }\href {\doibase
  10.1007/s00216-018-1034-6} {\bibfield  {journal} {\bibinfo  {journal} {Anal.
  Bioanal. Cham.}\ }\textbf {\bibinfo {volume} {410}},\ \bibinfo {pages} {3385}
  (\bibinfo {year} {2018})}\BibitemShut {NoStop}%
\bibitem [{\citenamefont {Fornell}\ \emph {et~al.}(2018)\citenamefont
  {Fornell}, \citenamefont {Cushing}, \citenamefont {Nilsson},\ and\
  \citenamefont {Tenje}}]{Fornell2018}%
  \BibitemOpen
  \bibfield  {author} {\bibinfo {author} {\bibfnamefont {A.}~\bibnamefont
  {Fornell}}, \bibinfo {author} {\bibfnamefont {K.}~\bibnamefont {Cushing}},
  \bibinfo {author} {\bibfnamefont {J.}~\bibnamefont {Nilsson}}, \ and\
  \bibinfo {author} {\bibfnamefont {M.}~\bibnamefont {Tenje}},\ }\bibfield
  {title} {\bibinfo {title} {Binary particle separation in droplet
  microfluidics using acoustophoresis},\ }\href {\doibase 10.1063/1.5020356}
  {\bibfield  {journal} {\bibinfo  {journal} {Appl. Phys. Lett.}\ }\textbf
  {\bibinfo {volume} {112}},\ \bibinfo {pages} {063701} (\bibinfo {year}
  {2018})}\BibitemShut {NoStop}%
\bibitem [{\citenamefont {Petersson}\ \emph {et~al.}(2018)\citenamefont
  {Petersson}, \citenamefont {Jakobsson}, \citenamefont {Ohlsson},
  \citenamefont {Augustsson}, \citenamefont {Scheding}, \citenamefont {Malm},\
  and\ \citenamefont {Laurell}}]{Petersson2018}%
  \BibitemOpen
  \bibfield  {author} {\bibinfo {author} {\bibfnamefont {K.}~\bibnamefont
  {Petersson}}, \bibinfo {author} {\bibfnamefont {O.}~\bibnamefont
  {Jakobsson}}, \bibinfo {author} {\bibfnamefont {P.}~\bibnamefont {Ohlsson}},
  \bibinfo {author} {\bibfnamefont {P.}~\bibnamefont {Augustsson}}, \bibinfo
  {author} {\bibfnamefont {S.}~\bibnamefont {Scheding}}, \bibinfo {author}
  {\bibfnamefont {J.}~\bibnamefont {Malm}}, \ and\ \bibinfo {author}
  {\bibfnamefont {T.}~\bibnamefont {Laurell}},\ }\bibfield  {title} {\bibinfo
  {title} {Acoustofluidic hematocrit determination},\ }\href {\doibase
  10.1016/j.aca.2017.11.037} {\bibfield  {journal} {\bibinfo  {journal} {Anal.
  Chim. Acta}\ }\textbf {\bibinfo {volume} {1000}},\ \bibinfo {pages} {199}
  (\bibinfo {year} {2018})}\BibitemShut {NoStop}%
\bibitem [{\citenamefont {Magnusson}\ \emph {et~al.}(2017)\citenamefont
  {Magnusson}, \citenamefont {Augustsson}, \citenamefont {Lenshof},
  \citenamefont {Ceder}, \citenamefont {Laurell},\ and\ \citenamefont
  {Lilja}}]{Magnusson2017}%
  \BibitemOpen
  \bibfield  {author} {\bibinfo {author} {\bibfnamefont {C.}~\bibnamefont
  {Magnusson}}, \bibinfo {author} {\bibfnamefont {P.}~\bibnamefont
  {Augustsson}}, \bibinfo {author} {\bibfnamefont {A.}~\bibnamefont {Lenshof}},
  \bibinfo {author} {\bibfnamefont {Y.}~\bibnamefont {Ceder}}, \bibinfo
  {author} {\bibfnamefont {T.}~\bibnamefont {Laurell}}, \ and\ \bibinfo
  {author} {\bibfnamefont {H.}~\bibnamefont {Lilja}},\ }\bibfield  {title}
  {\bibinfo {title} {Clinical-scale cell-surface-marker independent acoustic
  microfluidic enrichment of tumor cells from blood},\ }\href {\doibase
  10.1021/acs.analchem.7b01458} {\bibfield  {journal} {\bibinfo  {journal}
  {Anal. Chem.}\ }\textbf {\bibinfo {volume} {89}},\ \bibinfo {pages} {11954}
  (\bibinfo {year} {2017})}\BibitemShut {NoStop}%
\bibitem [{\citenamefont {Ahmed}\ \emph {et~al.}(2016)\citenamefont {Ahmed},
  \citenamefont {Ozcelik}, \citenamefont {Bojanala}, \citenamefont {Nama},
  \citenamefont {Upadhyay}, \citenamefont {Chen}, \citenamefont {Hanna-Rose},\
  and\ \citenamefont {Huang}}]{Ahmed2016}%
  \BibitemOpen
  \bibfield  {author} {\bibinfo {author} {\bibfnamefont {D.}~\bibnamefont
  {Ahmed}}, \bibinfo {author} {\bibfnamefont {A.}~\bibnamefont {Ozcelik}},
  \bibinfo {author} {\bibfnamefont {N.}~\bibnamefont {Bojanala}}, \bibinfo
  {author} {\bibfnamefont {N.}~\bibnamefont {Nama}}, \bibinfo {author}
  {\bibfnamefont {A.}~\bibnamefont {Upadhyay}}, \bibinfo {author}
  {\bibfnamefont {Y.}~\bibnamefont {Chen}}, \bibinfo {author} {\bibfnamefont
  {W.}~\bibnamefont {Hanna-Rose}}, \ and\ \bibinfo {author} {\bibfnamefont
  {T.~J.}\ \bibnamefont {Huang}},\ }\bibfield  {title} {\bibinfo {title}
  {Rotational manipulation of single cells and organisms using acoustic
  waves},\ }\href {\doibase 10.1038/ncomms11085} {\bibfield  {journal}
  {\bibinfo  {journal} {Nat. Commun.}\ }\textbf {\bibinfo {volume} {7}},\
  \bibinfo {pages} {11085} (\bibinfo {year} {2016})}\BibitemShut {NoStop}%
\bibitem [{\citenamefont {Zhou}\ \emph {et~al.}(2017)\citenamefont {Zhou},
  \citenamefont {Wang}, \citenamefont {Wang}, \citenamefont {Huang},
  \citenamefont {Niu}, \citenamefont {Li}, \citenamefont {Cai}, \citenamefont
  {Chen}, \citenamefont {Liu}, \citenamefont {Zhang}, \citenamefont {Cheng},
  \citenamefont {Kang}, \citenamefont {Meng},\ and\ \citenamefont
  {Zheng}}]{Zhou2017}%
  \BibitemOpen
  \bibfield  {author} {\bibinfo {author} {\bibfnamefont {W.}~\bibnamefont
  {Zhou}}, \bibinfo {author} {\bibfnamefont {J.}~\bibnamefont {Wang}}, \bibinfo
  {author} {\bibfnamefont {K.}~\bibnamefont {Wang}}, \bibinfo {author}
  {\bibfnamefont {B.}~\bibnamefont {Huang}}, \bibinfo {author} {\bibfnamefont
  {L.}~\bibnamefont {Niu}}, \bibinfo {author} {\bibfnamefont {F.}~\bibnamefont
  {Li}}, \bibinfo {author} {\bibfnamefont {F.}~\bibnamefont {Cai}}, \bibinfo
  {author} {\bibfnamefont {Y.}~\bibnamefont {Chen}}, \bibinfo {author}
  {\bibfnamefont {X.}~\bibnamefont {Liu}}, \bibinfo {author} {\bibfnamefont
  {X.}~\bibnamefont {Zhang}}, \bibinfo {author} {\bibfnamefont
  {H.}~\bibnamefont {Cheng}}, \bibinfo {author} {\bibfnamefont
  {L.}~\bibnamefont {Kang}}, \bibinfo {author} {\bibfnamefont {L.}~\bibnamefont
  {Meng}}, \ and\ \bibinfo {author} {\bibfnamefont {H.}~\bibnamefont {Zheng}},\
  }\bibfield  {title} {\bibinfo {title} {Ultrasound neuro-modulation chip:
  activation of sensory neurons in caenorhabditis elegans by surface acoustic
  waves},\ }\href {\doibase 10.1039/c7lc00163k} {\bibfield  {journal} {\bibinfo
   {journal} {Lab Chip}\ }\textbf {\bibinfo {volume} {17}},\ \bibinfo {pages}
  {1725} (\bibinfo {year} {2017})}\BibitemShut {NoStop}%
\bibitem [{\citenamefont {Sehgal}\ and\ \citenamefont
  {Kirby}(2017)}]{Sehgal2017}%
  \BibitemOpen
  \bibfield  {author} {\bibinfo {author} {\bibfnamefont {P.}~\bibnamefont
  {Sehgal}}\ and\ \bibinfo {author} {\bibfnamefont {B.~J.}\ \bibnamefont
  {Kirby}},\ }\bibfield  {title} {\bibinfo {title} {Separation of 300 and 100
  nm particles in fabry-perot acoustofluidic resonators},\ }\href {\doibase
  10.1021/acs.analchem.7b02858} {\bibfield  {journal} {\bibinfo  {journal}
  {Anal. Chem.}\ }\textbf {\bibinfo {volume} {89}},\ \bibinfo {pages} {12192}
  (\bibinfo {year} {2017})}\BibitemShut {NoStop}%
\bibitem [{\citenamefont {Wu}\ \emph {et~al.}(2017)\citenamefont {Wu},
  \citenamefont {Mao}, \citenamefont {Chen}, \citenamefont {Bachman},
  \citenamefont {Chen}, \citenamefont {Rufo}, \citenamefont {Ren},
  \citenamefont {Li}, \citenamefont {Wang},\ and\ \citenamefont
  {Huang}}]{Wu2017}%
  \BibitemOpen
  \bibfield  {author} {\bibinfo {author} {\bibfnamefont {M.}~\bibnamefont
  {Wu}}, \bibinfo {author} {\bibfnamefont {Z.}~\bibnamefont {Mao}}, \bibinfo
  {author} {\bibfnamefont {K.}~\bibnamefont {Chen}}, \bibinfo {author}
  {\bibfnamefont {H.}~\bibnamefont {Bachman}}, \bibinfo {author} {\bibfnamefont
  {Y.}~\bibnamefont {Chen}}, \bibinfo {author} {\bibfnamefont {J.}~\bibnamefont
  {Rufo}}, \bibinfo {author} {\bibfnamefont {L.}~\bibnamefont {Ren}}, \bibinfo
  {author} {\bibfnamefont {P.}~\bibnamefont {Li}}, \bibinfo {author}
  {\bibfnamefont {L.}~\bibnamefont {Wang}}, \ and\ \bibinfo {author}
  {\bibfnamefont {T.~J.}\ \bibnamefont {Huang}},\ }\bibfield  {title} {\bibinfo
  {title} {Acoustic separation of nanoparticles in continuous flow},\ }\href
  {\doibase 10.1002/adfm.201606039} {\bibfield  {journal} {\bibinfo  {journal}
  {Adv. Funct. Mater.}\ }\textbf {\bibinfo {volume} {27}},\ \bibinfo {pages}
  {1606039} (\bibinfo {year} {2017})}\BibitemShut {NoStop}%
\bibitem [{\citenamefont {Collins}\ \emph {et~al.}(2018)\citenamefont
  {Collins}, \citenamefont {O'Rorke}, \citenamefont {Devendran}, \citenamefont
  {Ma}, \citenamefont {Han}, \citenamefont {Neild},\ and\ \citenamefont
  {Ai}}]{Collins2018}%
  \BibitemOpen
  \bibfield  {author} {\bibinfo {author} {\bibfnamefont {D.~J.}\ \bibnamefont
  {Collins}}, \bibinfo {author} {\bibfnamefont {R.}~\bibnamefont {O'Rorke}},
  \bibinfo {author} {\bibfnamefont {C.}~\bibnamefont {Devendran}}, \bibinfo
  {author} {\bibfnamefont {Z.}~\bibnamefont {Ma}}, \bibinfo {author}
  {\bibfnamefont {J.}~\bibnamefont {Han}}, \bibinfo {author} {\bibfnamefont
  {A.}~\bibnamefont {Neild}}, \ and\ \bibinfo {author} {\bibfnamefont
  {Y.}~\bibnamefont {Ai}},\ }\bibfield  {title} {\bibinfo {title} {Self-aligned
  acoustofluidic particle focusing and patterning in microfluidic channels from
  channel-based acoustic waveguides},\ }\href {\doibase
  10.1103/PhysRevLett.120.074502} {\bibfield  {journal} {\bibinfo  {journal}
  {Phys. Rev. Lett.}\ }\textbf {\bibinfo {volume} {120}},\ \bibinfo {pages}
  {074502} (\bibinfo {year} {2018})}\BibitemShut {NoStop}%
\bibitem [{\citenamefont {Ung}\ \emph {et~al.}(2017)\citenamefont {Ung},
  \citenamefont {Mutafopulos}, \citenamefont {Spink}, \citenamefont {Rambach},
  \citenamefont {Franke},\ and\ \citenamefont {Weitz}}]{Ung2017}%
  \BibitemOpen
  \bibfield  {author} {\bibinfo {author} {\bibfnamefont {W.~L.}\ \bibnamefont
  {Ung}}, \bibinfo {author} {\bibfnamefont {K.}~\bibnamefont {Mutafopulos}},
  \bibinfo {author} {\bibfnamefont {P.}~\bibnamefont {Spink}}, \bibinfo
  {author} {\bibfnamefont {R.~W.}\ \bibnamefont {Rambach}}, \bibinfo {author}
  {\bibfnamefont {T.}~\bibnamefont {Franke}}, \ and\ \bibinfo {author}
  {\bibfnamefont {D.~A.}\ \bibnamefont {Weitz}},\ }\bibfield  {title} {\bibinfo
  {title} {Enhanced surface acoustic wave cell sorting by 3d microfluidic-chip
  design},\ }\href {\doibase 10.1039/c7lc00715a} {\bibfield  {journal}
  {\bibinfo  {journal} {Lab Chip}\ }\textbf {\bibinfo {volume} {17}},\ \bibinfo
  {pages} {4059} (\bibinfo {year} {2017})}\BibitemShut {NoStop}%
\bibitem [{\citenamefont {Park}\ \emph {et~al.}(2017)\citenamefont {Park},
  \citenamefont {Park}, \citenamefont {Jung}, \citenamefont {Ahmed},\ and\
  \citenamefont {Sung}}]{Park2017}%
  \BibitemOpen
  \bibfield  {author} {\bibinfo {author} {\bibfnamefont {K.}~\bibnamefont
  {Park}}, \bibinfo {author} {\bibfnamefont {J.}~\bibnamefont {Park}}, \bibinfo
  {author} {\bibfnamefont {J.~H.}\ \bibnamefont {Jung}}, \bibinfo {author}
  {\bibfnamefont {H.}~\bibnamefont {Ahmed}}, \ and\ \bibinfo {author}
  {\bibfnamefont {H.~J.}\ \bibnamefont {Sung}},\ }\bibfield  {title} {\bibinfo
  {title} {In-droplet microparticle separation using travelling surface
  acoustic wave},\ }\href {\doibase 10.1063/1.5010219} {\bibfield  {journal}
  {\bibinfo  {journal} {Biomicrofluidics}\ }\textbf {\bibinfo {volume} {11}},\
  \bibinfo {pages} {064112} (\bibinfo {year} {2017})}\BibitemShut {NoStop}%
\bibitem [{\citenamefont {Kim}\ and\ \citenamefont {Meng}(2016)}]{Kim2016b}%
  \BibitemOpen
  \bibfield  {author} {\bibinfo {author} {\bibfnamefont {B.~J.}\ \bibnamefont
  {Kim}}\ and\ \bibinfo {author} {\bibfnamefont {E.}~\bibnamefont {Meng}},\
  }\bibfield  {title} {\bibinfo {title} {Review of polymer {MEMS}
  micromachining},\ }\href@noop {} {\bibfield  {journal} {\bibinfo  {journal}
  {J. Micromech. Microeng.}\ }\textbf {\bibinfo {volume} {26}},\ \bibinfo
  {pages} {013001} (\bibinfo {year} {2016})}\BibitemShut {NoStop}%
\bibitem [{\citenamefont {Harris}\ \emph {et~al.}(2012)\citenamefont {Harris},
  \citenamefont {Hill}, \citenamefont {Keating},\ and\ \citenamefont
  {Baylac-Choulet}}]{Harris2012}%
  \BibitemOpen
  \bibfield  {author} {\bibinfo {author} {\bibfnamefont {N.}~\bibnamefont
  {Harris}}, \bibinfo {author} {\bibfnamefont {M.}~\bibnamefont {Hill}},
  \bibinfo {author} {\bibfnamefont {A.}~\bibnamefont {Keating}}, \ and\
  \bibinfo {author} {\bibfnamefont {P.}~\bibnamefont {Baylac-Choulet}},\
  }\bibfield  {title} {\bibinfo {title} {A lateral mode flow-through {PMMA}
  ultrasonic separator},\ }\href
  {http://www.ijabme.org/images/stories/ijabme/2012/ijabme-5-no-5-2012.pdf}
  {\bibfield  {journal} {\bibinfo  {journal} {Intl. J. Appl. Biomed. Eng.}\
  }\textbf {\bibinfo {volume} {5}},\ \bibinfo {pages} {20} (\bibinfo {year}
  {2012})}\BibitemShut {NoStop}%
\bibitem [{\citenamefont {Mueller}\ \emph {et~al.}(2013)\citenamefont
  {Mueller}, \citenamefont {Lever}, \citenamefont {Nguyen}, \citenamefont
  {Comolli},\ and\ \citenamefont {Fiering}}]{Mueller2013}%
  \BibitemOpen
  \bibfield  {author} {\bibinfo {author} {\bibfnamefont {A.}~\bibnamefont
  {Mueller}}, \bibinfo {author} {\bibfnamefont {A.}~\bibnamefont {Lever}},
  \bibinfo {author} {\bibfnamefont {T.~V.}\ \bibnamefont {Nguyen}}, \bibinfo
  {author} {\bibfnamefont {J.}~\bibnamefont {Comolli}}, \ and\ \bibinfo
  {author} {\bibfnamefont {J.}~\bibnamefont {Fiering}},\ }\bibfield  {title}
  {\bibinfo {title} {Continuous acoustic separation in a thermoplastic
  microchannel},\ }\href {\doibase 10.1088/0960-1317/23/12/125006} {\bibfield
  {journal} {\bibinfo  {journal} {J Micromech Microeng}\ }\textbf {\bibinfo
  {volume} {23}},\ \bibinfo {pages} {125006} (\bibinfo {year}
  {2013})}\BibitemShut {NoStop}%
\bibitem [{\citenamefont {Gonzalez}\ \emph {et~al.}(2015)\citenamefont
  {Gonzalez}, \citenamefont {Tijero}, \citenamefont {Martin}, \citenamefont
  {Acosta}, \citenamefont {Berganzo}, \citenamefont {Castillejo}, \citenamefont
  {Bouali},\ and\ \citenamefont {Luis~Soto}}]{Gonzalez2015}%
  \BibitemOpen
  \bibfield  {author} {\bibinfo {author} {\bibfnamefont {I.}~\bibnamefont
  {Gonzalez}}, \bibinfo {author} {\bibfnamefont {M.}~\bibnamefont {Tijero}},
  \bibinfo {author} {\bibfnamefont {A.}~\bibnamefont {Martin}}, \bibinfo
  {author} {\bibfnamefont {V.}~\bibnamefont {Acosta}}, \bibinfo {author}
  {\bibfnamefont {J.}~\bibnamefont {Berganzo}}, \bibinfo {author}
  {\bibfnamefont {A.}~\bibnamefont {Castillejo}}, \bibinfo {author}
  {\bibfnamefont {M.~M.}\ \bibnamefont {Bouali}}, \ and\ \bibinfo {author}
  {\bibfnamefont {J.}~\bibnamefont {Luis~Soto}},\ }\bibfield  {title} {\bibinfo
  {title} {Optimizing polymer lab-on-chip platforms for ultrasonic
  manipulation: Influence of the substrate},\ }\href {\doibase
  10.3390/mi6050574} {\bibfield  {journal} {\bibinfo  {journal}
  {Micromachines}\ }\textbf {\bibinfo {volume} {6}},\ \bibinfo {pages} {574}
  (\bibinfo {year} {2015})}\BibitemShut {NoStop}%
\bibitem [{\citenamefont {Yang}\ \emph {et~al.}()\citenamefont {Yang},
  \citenamefont {Li}, \citenamefont {Li}, \citenamefont {Shao}, \citenamefont
  {Bai},\ and\ \citenamefont {Cui}}]{Yang2017}%
  \BibitemOpen
  \bibfield  {author} {\bibinfo {author} {\bibfnamefont {C.}~\bibnamefont
  {Yang}}, \bibinfo {author} {\bibfnamefont {Z.}~\bibnamefont {Li}}, \bibinfo
  {author} {\bibfnamefont {P.}~\bibnamefont {Li}}, \bibinfo {author}
  {\bibfnamefont {W.}~\bibnamefont {Shao}}, \bibinfo {author} {\bibfnamefont
  {P.}~\bibnamefont {Bai}}, \ and\ \bibinfo {author} {\bibfnamefont
  {Y.}~\bibnamefont {Cui}},\ }\bibfield  {title} {\bibinfo {title} {Acoustic
  particle sorting by integrated micromachined ultrasound transducers on
  polymerbased microchips},\ }\href@noop {} {\bibinfo  {journal} {IEEE
  International Ultrasonics Symposium (IUS)}\ }\BibitemShut {NoStop}%
\bibitem [{\citenamefont {Savage}\ \emph {et~al.}(2017)\citenamefont {Savage},
  \citenamefont {Burns},\ and\ \citenamefont {Fiering}}]{Savage2017}%
  \BibitemOpen
\bibfield  {journal} {  }\bibfield  {author} {\bibinfo {author} {\bibfnamefont
  {W.~J.}\ \bibnamefont {Savage}}, \bibinfo {author} {\bibfnamefont {J.~R.}\
  \bibnamefont {Burns}}, \ and\ \bibinfo {author} {\bibfnamefont
  {J.}~\bibnamefont {Fiering}},\ }\bibfield  {title} {\bibinfo {title} {Safety
  of acoustic separation in plastic devices for extracorporeal blood
  processing},\ }\href {\doibase 10.1111/trf.14158} {\bibfield  {journal}
  {\bibinfo  {journal} {Transfusion}\ }\textbf {\bibinfo {volume} {57}},\
  \bibinfo {pages} {1818} (\bibinfo {year} {2017})}\BibitemShut {NoStop}%
\bibitem [{\citenamefont {Silva}\ \emph {et~al.}(2017)\citenamefont {Silva},
  \citenamefont {Dow}, \citenamefont {Dubay}, \citenamefont {Lissandrello},
  \citenamefont {Holder}, \citenamefont {Densmore},\ and\ \citenamefont
  {Fiering}}]{Silva2017}%
  \BibitemOpen
  \bibfield  {author} {\bibinfo {author} {\bibfnamefont {R.}~\bibnamefont
  {Silva}}, \bibinfo {author} {\bibfnamefont {P.}~\bibnamefont {Dow}}, \bibinfo
  {author} {\bibfnamefont {R.}~\bibnamefont {Dubay}}, \bibinfo {author}
  {\bibfnamefont {C.}~\bibnamefont {Lissandrello}}, \bibinfo {author}
  {\bibfnamefont {J.}~\bibnamefont {Holder}}, \bibinfo {author} {\bibfnamefont
  {D.}~\bibnamefont {Densmore}}, \ and\ \bibinfo {author} {\bibfnamefont
  {J.}~\bibnamefont {Fiering}},\ }\bibfield  {title} {\bibinfo {title} {Rapid
  prototyping and parametric optimization of plastic acoustofluidic devices for
  blood-bacteria separation},\ }\href {\doibase 10.1007/s10544-017-0210-3}
  {\bibfield  {journal} {\bibinfo  {journal} {Biomedical Microdevices}\
  }\textbf {\bibinfo {volume} {19}},\ \bibinfo {pages} {70} (\bibinfo {year}
  {2017})}\BibitemShut {NoStop}%
\bibitem [{\citenamefont {Lissandrello}\ \emph {et~al.}(2018)\citenamefont
  {Lissandrello}, \citenamefont {Dubay}, \citenamefont {Kotz},\ and\
  \citenamefont {Fiering}}]{Lissandrello2018}%
  \BibitemOpen
  \bibfield  {author} {\bibinfo {author} {\bibfnamefont {C.}~\bibnamefont
  {Lissandrello}}, \bibinfo {author} {\bibfnamefont {R.}~\bibnamefont {Dubay}},
  \bibinfo {author} {\bibfnamefont {K.~T.}\ \bibnamefont {Kotz}}, \ and\
  \bibinfo {author} {\bibfnamefont {J.}~\bibnamefont {Fiering}},\ }\bibfield
  {title} {\bibinfo {title} {Purification of lymphocytes by acoustic separation
  in plastic microchannels},\ }\href {\doibase 10.1177/2472630317749944}
  {\bibfield  {journal} {\bibinfo  {journal} {{SLAS} Technology}\ }\textbf
  {\bibinfo {volume} {23}},\ \bibinfo {pages} {352} (\bibinfo {year}
  {2018})}\BibitemShut {NoStop}%
\bibitem [{\citenamefont {Muller}\ \emph {et~al.}(2012)\citenamefont {Muller},
  \citenamefont {Barnkob}, \citenamefont {Jensen},\ and\ \citenamefont
  {Bruus}}]{Muller2012}%
  \BibitemOpen
  \bibfield  {author} {\bibinfo {author} {\bibfnamefont {P.~B.}\ \bibnamefont
  {Muller}}, \bibinfo {author} {\bibfnamefont {R.}~\bibnamefont {Barnkob}},
  \bibinfo {author} {\bibfnamefont {M.~J.~H.}\ \bibnamefont {Jensen}}, \ and\
  \bibinfo {author} {\bibfnamefont {H.}~\bibnamefont {Bruus}},\ }\bibfield
  {title} {\bibinfo {title} {A numerical study of microparticle acoustophoresis
  driven by acoustic radiation forces and streaming-induced drag forces},\
  }\href {\doibase 10.1039/C2LC40612H} {\bibfield  {journal} {\bibinfo
  {journal} {Lab Chip}\ }\textbf {\bibinfo {volume} {12}},\ \bibinfo {pages}
  {4617} (\bibinfo {year} {2012})}\BibitemShut {NoStop}%
\bibitem [{\citenamefont {Muller}\ and\ \citenamefont
  {Bruus}(2014)}]{Muller2014}%
  \BibitemOpen
  \bibfield  {author} {\bibinfo {author} {\bibfnamefont {P.~B.}\ \bibnamefont
  {Muller}}\ and\ \bibinfo {author} {\bibfnamefont {H.}~\bibnamefont {Bruus}},\
  }\bibfield  {title} {\bibinfo {title} {Numerical study of thermoviscous
  effects in ultrasound-induced acoustic streaming in microchannels},\ }\href
  {\doibase 10.1103/PhysRevE.90.043016} {\bibfield  {journal} {\bibinfo
  {journal} {Phys. Rev. E}\ }\textbf {\bibinfo {volume} {90}},\ \bibinfo
  {pages} {043016} (\bibinfo {year} {2014})}\BibitemShut {NoStop}%
\bibitem [{\citenamefont {Muller}\ and\ \citenamefont
  {Bruus}(2015)}]{Muller2015}%
  \BibitemOpen
  \bibfield  {author} {\bibinfo {author} {\bibfnamefont {P.~B.}\ \bibnamefont
  {Muller}}\ and\ \bibinfo {author} {\bibfnamefont {H.}~\bibnamefont {Bruus}},\
  }\bibfield  {title} {\bibinfo {title} {Theoretical study of time-dependent,
  ultrasound-induced acoustic streaming in microchannels},\ }\href {\doibase
  10.1103/PhysRevE.92.063018} {\bibfield  {journal} {\bibinfo  {journal} {Phys.
  Rev. E}\ }\textbf {\bibinfo {volume} {92}},\ \bibinfo {pages} {063018}
  (\bibinfo {year} {2015})}\BibitemShut {NoStop}%
\bibitem [{\citenamefont {Bach}\ and\ \citenamefont {Bruus}(2018)}]{Bach2018}%
  \BibitemOpen
  \bibfield  {author} {\bibinfo {author} {\bibfnamefont {J.~S.}\ \bibnamefont
  {Bach}}\ and\ \bibinfo {author} {\bibfnamefont {H.}~\bibnamefont {Bruus}},\
  }\bibfield  {title} {\bibinfo {title} {Theory of pressure acoustics with
  viscous boundary layers and streaming in curved elastic cavities},\ }\href
  {\doibase 10.1121/1.5049579} {\bibfield  {journal} {\bibinfo  {journal} {J.
  Acoust. Soc. Am.}\ }\textbf {\bibinfo {volume} {144}},\ \bibinfo {pages}
  {766} (\bibinfo {year} {2018})}\BibitemShut {NoStop}%
\bibitem [{\citenamefont {Barnkob}\ \emph {et~al.}(2010)\citenamefont
  {Barnkob}, \citenamefont {Augustsson}, \citenamefont {Laurell},\ and\
  \citenamefont {Bruus}}]{Barnkob2010}%
  \BibitemOpen
  \bibfield  {author} {\bibinfo {author} {\bibfnamefont {R.}~\bibnamefont
  {Barnkob}}, \bibinfo {author} {\bibfnamefont {P.}~\bibnamefont {Augustsson}},
  \bibinfo {author} {\bibfnamefont {T.}~\bibnamefont {Laurell}}, \ and\
  \bibinfo {author} {\bibfnamefont {H.}~\bibnamefont {Bruus}},\ }\bibfield
  {title} {\bibinfo {title} {Measuring the local pressure amplitude in
  microchannel acoustophoresis},\ }\href {\doibase 10.1039/b920376a} {\bibfield
   {journal} {\bibinfo  {journal} {Lab Chip}\ }\textbf {\bibinfo {volume}
  {10}},\ \bibinfo {pages} {563} (\bibinfo {year} {2010})}\BibitemShut
  {NoStop}%
\bibitem [{\citenamefont {Augustsson}\ \emph {et~al.}(2011)\citenamefont
  {Augustsson}, \citenamefont {Barnkob}, \citenamefont {Wereley}, \citenamefont
  {Bruus},\ and\ \citenamefont {Laurell}}]{Augustsson2011}%
  \BibitemOpen
  \bibfield  {author} {\bibinfo {author} {\bibfnamefont {P.}~\bibnamefont
  {Augustsson}}, \bibinfo {author} {\bibfnamefont {R.}~\bibnamefont {Barnkob}},
  \bibinfo {author} {\bibfnamefont {S.~T.}\ \bibnamefont {Wereley}}, \bibinfo
  {author} {\bibfnamefont {H.}~\bibnamefont {Bruus}}, \ and\ \bibinfo {author}
  {\bibfnamefont {T.}~\bibnamefont {Laurell}},\ }\bibfield  {title} {\bibinfo
  {title} {Automated and temperature-controlled micro-piv measurements enabling
  long-term-stable microchannel acoustophoresis characterization},\ }\href
  {\doibase 10.1039/c1lc20637k} {\bibfield  {journal} {\bibinfo  {journal} {Lab
  Chip}\ }\textbf {\bibinfo {volume} {11}},\ \bibinfo {pages} {4152} (\bibinfo
  {year} {2011})}\BibitemShut {NoStop}%
\bibitem [{\citenamefont {Barnkob}\ \emph {et~al.}(2012)\citenamefont
  {Barnkob}, \citenamefont {Augustsson}, \citenamefont {Laurell},\ and\
  \citenamefont {Bruus}}]{Barnkob2012a}%
  \BibitemOpen
  \bibfield  {author} {\bibinfo {author} {\bibfnamefont {R.}~\bibnamefont
  {Barnkob}}, \bibinfo {author} {\bibfnamefont {P.}~\bibnamefont {Augustsson}},
  \bibinfo {author} {\bibfnamefont {T.}~\bibnamefont {Laurell}}, \ and\
  \bibinfo {author} {\bibfnamefont {H.}~\bibnamefont {Bruus}},\ }\bibfield
  {title} {\bibinfo {title} {Acoustic radiation- and streaming-induced
  microparticle velocities determined by microparticle image velocimetry in an
  ultrasound symmetry plane},\ }\href {\doibase 10.1103/PhysRevE.86.056307}
  {\bibfield  {journal} {\bibinfo  {journal} {Phys. Rev. E}\ }\textbf {\bibinfo
  {volume} {86}},\ \bibinfo {pages} {056307} (\bibinfo {year}
  {2012})}\BibitemShut {NoStop}%
\bibitem [{\citenamefont {Muller}\ \emph {et~al.}(2013)\citenamefont {Muller},
  \citenamefont {Rossi}, \citenamefont {Marin}, \citenamefont {Barnkob},
  \citenamefont {Augustsson}, \citenamefont {Laurell}, \citenamefont
  {K\"{a}hler},\ and\ \citenamefont {Bruus}}]{Muller2013}%
  \BibitemOpen
  \bibfield  {author} {\bibinfo {author} {\bibfnamefont {P.~B.}\ \bibnamefont
  {Muller}}, \bibinfo {author} {\bibfnamefont {M.}~\bibnamefont {Rossi}},
  \bibinfo {author} {\bibfnamefont {A.~G.}\ \bibnamefont {Marin}}, \bibinfo
  {author} {\bibfnamefont {R.}~\bibnamefont {Barnkob}}, \bibinfo {author}
  {\bibfnamefont {P.}~\bibnamefont {Augustsson}}, \bibinfo {author}
  {\bibfnamefont {T.}~\bibnamefont {Laurell}}, \bibinfo {author} {\bibfnamefont
  {C.~J.}\ \bibnamefont {K\"{a}hler}}, \ and\ \bibinfo {author} {\bibfnamefont
  {H.}~\bibnamefont {Bruus}},\ }\bibfield  {title} {\bibinfo {title}
  {Ultrasound-induced acoustophoretic motion of microparticles in three
  dimensions},\ }\href {\doibase 10.1103/PhysRevE.88.023006} {\bibfield
  {journal} {\bibinfo  {journal} {Phys. Rev. E}\ }\textbf {\bibinfo {volume}
  {88}},\ \bibinfo {pages} {023006} (\bibinfo {year} {2013})}\BibitemShut
  {NoStop}%
\bibitem [{\citenamefont {Sutherland}\ and\ \citenamefont
  {Lingle}(1972)}]{Sutherland1972}%
  \BibitemOpen
  \bibfield  {author} {\bibinfo {author} {\bibfnamefont {H.}~\bibnamefont
  {Sutherland}}\ and\ \bibinfo {author} {\bibfnamefont {R.}~\bibnamefont
  {Lingle}},\ }\bibfield  {title} {\bibinfo {title} {Acoustic characterization
  of polymethyl methacrylate and 3 epoxy formulations},\ }\href {\doibase
  10.1063/1.1660868} {\bibfield  {journal} {\bibinfo  {journal} {J. Appl.
  Phys.}\ }\textbf {\bibinfo {volume} {43}},\ \bibinfo {pages} {4022} (\bibinfo
  {year} {1972})}\BibitemShut {NoStop}%
\bibitem [{\citenamefont {Sutherland}(1978)}]{Sutherland1978}%
  \BibitemOpen
  \bibfield  {author} {\bibinfo {author} {\bibfnamefont {H.}~\bibnamefont
  {Sutherland}},\ }\bibfield  {title} {\bibinfo {title} {Acoustical
  determination of shear relaxation functions for polymethyl methacrylate and
  {Epon 828-Z}},\ }\href {\doibase 10.1063/1.325403} {\bibfield  {journal}
  {\bibinfo  {journal} {J. Appl. Phys.}\ }\textbf {\bibinfo {volume} {49}},\
  \bibinfo {pages} {3941} (\bibinfo {year} {1978})}\BibitemShut {NoStop}%
\bibitem [{\citenamefont {Carlson}\ \emph {et~al.}(2003)\citenamefont
  {Carlson}, \citenamefont {van Deventer}, \citenamefont {Scolan},\ and\
  \citenamefont {Carlander}}]{Carlson2003}%
  \BibitemOpen
  \bibfield  {author} {\bibinfo {author} {\bibfnamefont {J.~E.}\ \bibnamefont
  {Carlson}}, \bibinfo {author} {\bibfnamefont {J.}~\bibnamefont {van
  Deventer}}, \bibinfo {author} {\bibfnamefont {A.}~\bibnamefont {Scolan}}, \
  and\ \bibinfo {author} {\bibfnamefont {C.}~\bibnamefont {Carlander}},\
  }\bibfield  {title} {\bibinfo {title} {Frequency and temperature dependence
  of acoustic properties of polymers used in pulse-echo systems},\ }\href
  {\doibase 10.1109/ULTSYM.2003.1293541} {\bibfield  {journal} {\bibinfo
  {journal} {2003 IEEE Symposium on Ultrasonics}\ ,\ \bibinfo {pages} {885}}
  (\bibinfo {year} {2003})}\BibitemShut {NoStop}%
\bibitem [{\citenamefont {Mott}\ \emph {et~al.}(2008)\citenamefont {Mott},
  \citenamefont {Dorgan},\ and\ \citenamefont {Roland}}]{Mott2008}%
  \BibitemOpen
  \bibfield  {author} {\bibinfo {author} {\bibfnamefont {P.~H.}\ \bibnamefont
  {Mott}}, \bibinfo {author} {\bibfnamefont {J.~R.}\ \bibnamefont {Dorgan}}, \
  and\ \bibinfo {author} {\bibfnamefont {C.~M.}\ \bibnamefont {Roland}},\
  }\bibfield  {title} {\bibinfo {title} {The bulk modulus and poisson's ratio
  of {''}incompressible{''} materials},\ }\href {\doibase
  10.1016/j.jsv.2008.01.026} {\bibfield  {journal} {\bibinfo  {journal} {J
  Sound Vibr}\ }\textbf {\bibinfo {volume} {312}},\ \bibinfo {pages} {572}
  (\bibinfo {year} {2008})}\BibitemShut {NoStop}%
\bibitem [{\citenamefont {Adler}(2014)}]{Adler2014}%
  \BibitemOpen
  \bibfield  {author} {\bibinfo {author} {\bibfnamefont {P.}~\bibnamefont
  {Adler}},\ }\bibinfo {title} {Novel materials and designs for cell separation
  by acoustofluidics},\ in\ \href@noop {} {\emph {\bibinfo {booktitle}
  {Master's thesis}}}\ (\bibinfo {year} {2014})\BibitemShut {NoStop}%
\bibitem [{\citenamefont {Kasarova}\ \emph {et~al.}(2015)\citenamefont
  {Kasarova}, \citenamefont {Sultanova},\ and\ \citenamefont
  {Nikolov}}]{Kasarova2015}%
  \BibitemOpen
  \bibfield  {author} {\bibinfo {author} {\bibfnamefont {S.}~\bibnamefont
  {Kasarova}}, \bibinfo {author} {\bibfnamefont {N.}~\bibnamefont {Sultanova}},
  \ and\ \bibinfo {author} {\bibfnamefont {I.}~\bibnamefont {Nikolov}},\
  }\bibfield  {title} {\bibinfo {title} {Polymer materials in optical design},\
  }\href@noop {} {\bibfield  {journal} {\bibinfo  {journal} {Bulgarian Chemical
  Communications}\ }\textbf {\bibinfo {volume} {47}},\ \bibinfo {pages} {44}
  (\bibinfo {year} {2015})}\BibitemShut {NoStop}%
\bibitem [{Cor()}]{Corning_Pyrex}%
  \BibitemOpen
  \href {http://www.valleydesign.com/Datasheets/Corning\%20Pyrex\%207740.pdf}
  {\emph {\bibinfo {title} {Glass Silicon Constraint Substrates}}},\ \bibinfo
  {organization} {CORNING},\ \bibinfo {address} {Houghton Park C-8, Corning, NY
  14831, USA},\ \bibinfo {note}
  {\url{http://www.valleydesign.com/Datasheets/Corning\%20Pyrex\%207740.pdf},
  accessed 11 November 2016}\BibitemShut {NoStop}%
\bibitem [{\citenamefont {Hopcroft}\ \emph {et~al.}(2010)\citenamefont
  {Hopcroft}, \citenamefont {Nix},\ and\ \citenamefont {Kenny}}]{Hopcroft2010}%
  \BibitemOpen
  \bibfield  {author} {\bibinfo {author} {\bibfnamefont {M.~A.}\ \bibnamefont
  {Hopcroft}}, \bibinfo {author} {\bibfnamefont {W.~D.}\ \bibnamefont {Nix}}, \
  and\ \bibinfo {author} {\bibfnamefont {T.~W.}\ \bibnamefont {Kenny}},\
  }\bibfield  {title} {\bibinfo {title} {What is the {Y}oung's modulus of
  silicon},\ }\href {\doibase 10.1109/JMEMS.2009.2039697} {\bibfield  {journal}
  {\bibinfo  {journal} {J. Microelectromech. Syst}\ }\textbf {\bibinfo {volume}
  {19}},\ \bibinfo {pages} {229} (\bibinfo {year} {2010})}\BibitemShut
  {NoStop}%
\bibitem [{\citenamefont {Hahn}\ and\ \citenamefont {Dual}(2015)}]{Hahn2015}%
  \BibitemOpen
  \bibfield  {author} {\bibinfo {author} {\bibfnamefont {P.}~\bibnamefont
  {Hahn}}\ and\ \bibinfo {author} {\bibfnamefont {J.}~\bibnamefont {Dual}},\
  }\bibfield  {title} {\bibinfo {title} {A numerically efficient damping model
  for acoustic resonances in microfluidic cavities},\ }\href {\doibase
  10.1063/1.4922986} {\bibfield  {journal} {\bibinfo  {journal} {Phys. Fluids}\
  }\textbf {\bibinfo {volume} {27}},\ \bibinfo {pages} {062005} (\bibinfo
  {year} {2015})}\BibitemShut {NoStop}%
\bibitem [{\citenamefont {Hartmann}\ and\ \citenamefont
  {Jarzynski}(1972)}]{Hartmann1972}%
  \BibitemOpen
  \bibfield  {author} {\bibinfo {author} {\bibfnamefont {B.}~\bibnamefont
  {Hartmann}}\ and\ \bibinfo {author} {\bibfnamefont {J.}~\bibnamefont
  {Jarzynski}},\ }\bibfield  {title} {\bibinfo {title} {Polymer sound speeds
  and elastic constants},\ }\href@noop {} {\bibfield  {journal} {\bibinfo
  {journal} {Naval Ordnance Laboratory, Report NOLTR}\ }\textbf {\bibinfo
  {volume} {72-269}},\ \bibinfo {pages} {1} (\bibinfo {year} {1972})},\
  \bibinfo {note} {\url{http://www.dtic.mil/dtic/tr/fulltext/u2/755695.pdf},
  accessed 16 August 2018}\BibitemShut {NoStop}%
\bibitem [{\citenamefont {Karlsen}\ and\ \citenamefont
  {Bruus}(2015)}]{Karlsen2015}%
  \BibitemOpen
  \bibfield  {author} {\bibinfo {author} {\bibfnamefont {J.~T.}\ \bibnamefont
  {Karlsen}}\ and\ \bibinfo {author} {\bibfnamefont {H.}~\bibnamefont
  {Bruus}},\ }\bibfield  {title} {\bibinfo {title} {Forces acting on a small
  particle in an acoustical field in a thermoviscous fluid},\ }\href {\doibase
  10.1103/PhysRevE.92.043010} {\bibfield  {journal} {\bibinfo  {journal} {Phys.
  Rev. E}\ }\textbf {\bibinfo {volume} {92}},\ \bibinfo {pages} {043010}
  (\bibinfo {year} {2015})}\BibitemShut {NoStop}%
\bibitem [{\citenamefont {Landau}\ and\ \citenamefont
  {Lifshitz}(1986)}]{Landau1986}%
  \BibitemOpen
  \bibfield  {author} {\bibinfo {author} {\bibfnamefont {L.~D.}\ \bibnamefont
  {Landau}}\ and\ \bibinfo {author} {\bibfnamefont {E.~M.}\ \bibnamefont
  {Lifshitz}},\ }\href@noop {} {\emph {\bibinfo {title} {Theory of Elasticity.
  Course of Theoretical Physics}}},\ \bibinfo {edition} {3rd}\ ed.,\
  Vol.~\bibinfo {volume} {7}\ (\bibinfo  {publisher} {Pergamon Press},\
  \bibinfo {address} {Oxford},\ \bibinfo {year} {1986})\BibitemShut {NoStop}%
\bibitem [{\citenamefont {Ley}\ and\ \citenamefont {Bruus}(2017)}]{Ley2017}%
  \BibitemOpen
  \bibfield  {author} {\bibinfo {author} {\bibfnamefont {M.~W.~H.}\
  \bibnamefont {Ley}}\ and\ \bibinfo {author} {\bibfnamefont {H.}~\bibnamefont
  {Bruus}},\ }\bibfield  {title} {\bibinfo {title} {Three-dimensional numerical
  modeling of acoustic trapping in glass capillaries},\ }\href {\doibase
  10.1103/PhysRevApplied.8.024020} {\bibfield  {journal} {\bibinfo  {journal}
  {Phys. Rev. Applied}\ }\textbf {\bibinfo {volume} {8}},\ \bibinfo {pages}
  {024020} (\bibinfo {year} {2017})}\BibitemShut {NoStop}%
\bibitem [{\citenamefont {Dual}\ and\ \citenamefont
  {Schwarz}(2012)}]{Dual2012}%
  \BibitemOpen
  \bibfield  {author} {\bibinfo {author} {\bibfnamefont {J.}~\bibnamefont
  {Dual}}\ and\ \bibinfo {author} {\bibfnamefont {T.}~\bibnamefont {Schwarz}},\
  }\bibfield  {title} {\bibinfo {title} {Acoustofluidics 3: Continuum mechanics
  for ultrasonic particle manipulation},\ }\href {\doibase 10.1039/C1LC20837C}
  {\bibfield  {journal} {\bibinfo  {journal} {Lab Chip}\ }\textbf {\bibinfo
  {volume} {12}},\ \bibinfo {pages} {244} (\bibinfo {year} {2012})}\BibitemShut
  {NoStop}%
\bibitem [{\citenamefont {Settnes}\ and\ \citenamefont
  {Bruus}(2012)}]{Settnes2012}%
  \BibitemOpen
  \bibfield  {author} {\bibinfo {author} {\bibfnamefont {M.}~\bibnamefont
  {Settnes}}\ and\ \bibinfo {author} {\bibfnamefont {H.}~\bibnamefont
  {Bruus}},\ }\bibfield  {title} {\bibinfo {title} {Forces acting on a small
  particle in an acoustical field in a viscous fluid},\ }\href {\doibase
  10.1103/PhysRevE.85.016327} {\bibfield  {journal} {\bibinfo  {journal} {Phys.
  Rev. E}\ }\textbf {\bibinfo {volume} {85}},\ \bibinfo {pages} {016327}
  (\bibinfo {year} {2012})}\BibitemShut {NoStop}%
\bibitem [{\citenamefont {{COMSOL Multiphysics 5.3a}}()}]{Comsol53a}%
  \BibitemOpen
  \bibfield  {author} {\bibinfo {author} {\bibnamefont {{COMSOL Multiphysics
  5.3a}}},\ }\href {www.comsol.com} {}\bibinfo {note} {\url{www.comsol.com}
  (2017)}\BibitemShut {NoStop}%
\bibitem [{\citenamefont {{AcouSort AB}}()}]{AcouSort}%
  \BibitemOpen
  \bibfield  {author} {\bibinfo {author} {\bibnamefont {{AcouSort AB}}},\
  }\href {www.acousort.com} {}\bibinfo {note} {\url{www.acousort.com}, private
  communication, 2018}\BibitemShut {NoStop}%
\end{thebibliography}

%

\end{document}